\documentclass[superscriptaddress,nofootinbib,a4paper,11pt]{article}
\usepackage{CJKutf8} 
\usepackage{jheppub}
\usepackage{graphicx}
\usepackage{subcaption}
\usepackage{float}
\usepackage{amsmath}
\usepackage{amssymb}
\usepackage[utf8]{inputenc}
\usepackage{hyperref}
\usepackage{color}
\usepackage{slashed}
\usepackage{ulem}
\allowdisplaybreaks

\newcommand{\be}{\begin{equation}}
\newcommand{\ee}{\end{equation}}

\definecolor{gesfblack}{rgb}{0,0,0}

\definecolor{gesfblue}{rgb}{0.08,0.42,0.76}

\definecolor{gesfgreen}{rgb}{0,1,0}

\definecolor{gesfgrey}{rgb}{0.5,0.5,0.5}

\definecolor{gesflanse}{rgb}{0.00,0.50,0.50}

\definecolor{gesfpurple}{rgb}{0.47,0.19,0.42}

\definecolor{gesfred}{rgb}{1,0,0}

\definecolor{gesfwhite}{rgb}{1,1,1}

\definecolor{gesfyellow}{rgb}{0.7,0.4,0.3}

\begin{document}

\title{\boldmath {Dark Parton Shower Effects for Cosmic Ray Boosted Dark Matter}}

\author[a]{Zirong Chen\begin{CJK}{UTF8}{gbsn}(陈子荣)\end{CJK},}
\author[b,c]{Shao-Feng Ge\begin{CJK}{UTF8}{gbsn}(葛韶锋)\end{CJK},}
\author[a]{Jinmian Li\begin{CJK}{UTF8}{gbsn}(李金勉)\end{CJK},}
\author[d]{Junle Pei\begin{CJK}{UTF8}{gbsn}(裴俊乐)\end{CJK},}
\author[a]{Feng Yang\begin{CJK}{UTF8}{gbsn}(杨峰)\end{CJK},}
\author[e,*]{Cong Zhang\begin{CJK}{UTF8}{gbsn}(张聪)\end{CJK}\note[*]{Corresponding author: zhangcong.phy@gmail.com}}

\affiliation[a]{College of Physics, Sichuan University, Chengdu 610065, China}
\affiliation[b]{State Key Laboratory of Dark Matter Physics, Tsung-Dao Lee Institute \& School of Physics and Astronomy, Shanghai Jiao Tong University, Shanghai 200240, China}
\affiliation[c]{Key Laboratory for Particle Astrophysics and Cosmology (MOE) \& Shanghai Key Laboratory for Particle Physics and Cosmology, Shanghai Jiao Tong University, Shanghai 200240, China}
\affiliation[d]{Institute of Physics, Henan Academy of Sciences, Zhengzhou 450046, China}
\affiliation[e]{Institute of Theoretical Physics, Chinese Academy of Sciences, Beijing, 100190, China}

\abstract{
We investigate the effects of dark parton showers on the direct detection of cosmic-ray boosted dark matter (CRDM), focusing on a dark-photon-mediated model with fermionic dark matter--electron interactions. Using a Monte Carlo framework to incorporate Sudakov form factors and kinematic dipole recoil schemes, we simulate the evolution of the CRDM energy spectrum under dark sector splitting. Our results reveal that dark parton showering induces an energy-dependent reshaping of the CRDM flux, particularly characterized by a depletion of the high-energy flux. For a dark matter {(DM)} mass of 1\,keV and a coupling of $g_D=3$, the CRDM flux can be enhanced by up to a factor of 1.12 in the $\mathcal{O}(10^{-2} \sim 1)$\,MeV energy range for $2m_\chi \lesssim m_{A^\prime} \lesssim 10^{-2}$\,MeV, whereas it is suppressed by more than 50\% at energies around 100\,MeV for $m_{A^\prime} \lesssim 10^{-3}$\,MeV. We then translate these effects into experimental sensitivities for PandaX-4T, Super-Kamiokande, and JUNO. For $m_{A^\prime} = 10^{-3}$\,MeV and $g_D=3$, the bounds on the kinetic mixing parameter $\epsilon^2$ are relaxed by factors of 1.02, 1.6, and 1.4, respectively. Incorporating the effects of DM parton distribution functions (PDFs) at neutrino detectors alongside dark parton shower effects further relaxes these bounds.

}


\maketitle

\section{Introduction}\label{sec:intro} 

Dark matter constitutes compelling evidence for
physics beyond the Standard Model (SM) \cite{Young:2016ala,Bertone:2016nfn,Arcadi:2017kky,Arbey:2021gdg,Cirelli:2024ssz}.
Extensive searches for potential DM signals have been
conducted across various avenues
\cite{Zhao:2020ezy,Cooley:2022ufh,Boveia:2022syt,Artuso:2022ouk},
including direct \cite{Cebrian:2021mvb,Cooley:2021rws,Aalbers:2022dzr,Misiaszek:2023sxe}, indirect \cite{Slatyer:2021qgc, Hutten:2022hud, deLaurentis:2022oqa},
and collider \cite{Ilten:2022lfq,Batell:2022dpx} searches.
In particular, direct detection experiments
aimed at observing nuclear or electron
recoils induced by DM scattering have reached the ton-scale
with PandaX-4T \cite{PandaX:2018wtu,PandaX-II:2020oim},
LZ \cite{LZ:2011rhn,LZ:2022lsv}, and XENONnT
\cite{XENON:2023cxc,XENON:2024wpa}.
However, these observations have yet to definitively
establish the nature of DM.
Due to the limited sensitivity of current detectors to energy
depositions below the keV scale, the direct detection of sub-GeV DM
poses significant challenges.

Current direct detection experiments primarily focus
on halo DM particles characterized by
non-relativistic
velocities of $v_{\chi} \sim 10^{{-3}} c$. A promising
avenue to probe
sub-GeV DM is by searching for these boosted components. 
Astrophysical processes, including primordial black hole
evaporation \cite{Carr:2020gox}, two-component
DM annihilation with mass hierarchy \cite{Agashe:2014yua}, and
DM semi-annihilation \cite{DEramo:2010keq}, among
others \cite{Jaeckel:2020oet,Herrera:2021puj}, can generate boosted DM populations within the Galactic halo.
These (near-)relativistic sub-components can produce
detectable signals in direct detection experiments,
even for DM masses well below 1\,GeV, potentially serving
as distinctive signatures for DM discovery. 
In particular, interactions between
DM and SM particles inevitably induce
scattering between high-energy cosmic rays (CRs) and DM.
These DM up-scattering processes generate a non-negligible
flux of boosted DM particles via either neutral neutrinos
\cite{Yin:2018yjn,Jho:2021rmn,Das:2021lcr}
or charged cosmic rays
\cite{Cappiello:2018hsu,Bringmann:2018cvk,Ema:2018bih}.
This mechanism allows even very light DM particles to deposit substantial energy in the detector
\cite{Bringmann:2018cvk,Wang:2019jtk,Ge:2020yuf,Cao:2020bwd,Jho:2020sku,Lei:2020mii,Flambaum:2020xxo,Dent:2019krz,Cho:2020mnc,Xia:2020apm,Feng:2021hyz,Wang:2021nbf,Emken:2021vmf,Bardhan:2022bdg,Alvey:2022pad,Xu:2024iny}.
The PROSPECT \cite{PROSPECT:2021awi},
PandaX \cite{PandaX-II:2021kai,PandaX:2024pme},
CDEX \cite{CDEX:2022fig},
Super-K \cite{Super-Kamiokande:2022ncz},
NEWSdm \cite{NEWSdm:2023qyb}, and
LZ \cite{LZ:2025iaw} collaborations
have analyzed their experimental data to search for such signals.
In particular, diurnal modulation
\cite{Ge:2020yuf,Fornal:2020npv,Chen:2021ifo,Xia:2021vbz}
and angular distribution \cite{Xia:2022tid}
analyses can help enhance signal sensitivity.

Large-volume neutrino experiments, such as Super-K~\cite{Super-Kamiokande:2002weg}, DUNE~\cite{DUNE:2020ypp}, and JUNO~\cite{JUNO:2021vlw}, situated deep underground, offer an alternative avenue for probing the boosted DM flux.
While neutrino detectors typically have higher energy thresholds for signal electrons compared to dedicated DM direct detection experiments and therefore reject a significant portion of DM scattering events, they compensate via their considerably larger detector volumes (typically tens to hundreds of kilotons).
This enables neutrino experiments to achieve competitive sensitivities to certain DM models, providing complementary information to traditional direct detection experiments~\cite{Ema:2018bih,Cappiello:2019qsw,Berger:2019ttc,Kim:2020ipj,Guo:2020drq,DeRoeck:2020ntj,Harnik:2020ugb,Ema:2020ulo,Dent:2020syp,Granelli:2022ysi,Berger:2022cab}.

Most existing studies have explored this possibility in a model-independent manner, assuming a constant scattering cross-section between DM and electrons or protons. However, in UV-complete models, DM interactions with SM particles are mediated by force carriers \cite{Cao:2020bwd, Jho:2020sku, Dent:2019krz, Cho:2020mnc, Feng:2021hyz, Bardhan:2022bdg, Alvey:2022pad, Xu:2024iny, Dent:2020syp}. 
In the context of CR-boosted DM, the energy scale of DM-SM particle scattering significantly exceeds the mass scales of the DM and mediator particles.
This hierarchy between energy scales, well-known in SM processes, leads to the emergence of large logarithmic contributions to the differential scattering cross-section in the presence of a light mediator.
Calculating these large logarithms requires resummation techniques, such as solving the Dokshitzer-Gribov-Lipatov-Altarelli-Parisi (DGLAP) evolution equation and performing parton shower simulations incorporating Sudakov form factors~\cite{Collins:1989bt}. The former effect has been studied in our previous work~\cite{Li:2022dqa}, where we included the DM Parton Distribution Function (PDF) to calculate the CRDM-electron scattering at neutrino detectors. Meanwhile, 
the spectrum of accelerated DM must be obtained by convolving the $\chi+e/p \to \chi+ e/p$ scattering process with subsequent parton showering.
While this parton showering effect is intimately linked to the underlying properties of the dark sector, its influence on observables has not yet been investigated in detail in the literature.

This work investigates the production
prospects of boosted CRDM in the context of a simplified electron-philic dark photon model with fermionic DM. Given a tiny kinetic mixing, the DM-dark photon
coupling dominates over the electron-dark photon coupling.
We first consider the CR up-scattering mechanism for DM acceleration.
Our analysis incorporates parton shower effects in the final state of boosted DM, accounting for both the subsequent decay and splitting of radiated dark photons into DM particles. We find that the dark parton shower alters the CRDM flux, with these changes subsequently reflected in the recoiling electron flux. Depending on the energy range, the differences in the recoil spectrum can reach tens of percent.
We further investigate the scattering of boosted DM with atomic electrons in the PandaX detector, targeting electron recoil energies of $\mathcal{O}(10)$\,keV. We also explore the detectability of boosted DM in several neutrino detectors, including Super-K and JUNO, focusing on electron recoil energies  $\gtrsim \mathcal{O}(10)$\,MeV. To provide a complete picture, the results presented in this study incorporate both the parton shower effects (which are its primary focus) and the DM PDF effects investigated in our previous work~\cite{Li:2022dqa}.

This paper is organized as follows. 
In Sec.~\ref{sec2}, we calculate the splitting functions with mass effects, which are crucial for the Monte Carlo simulation of final-state radiation (FSR).
In Sec.~\ref{sec3}, we derive the CRDM flux and develop the framework for simulating time-like parton showers. Sec.~\ref{sec4} presents the calculation of the boosted DM scattering cross section with either bound or free electrons. 
All relevant results are summarized and discussed in Sec.~\ref{sec5}.
Finally, we conclude our study in Sec.~\ref{sec6}.

\section{Dark Parton Shower}
\label{sec2}

The dark photon model \cite{Holdom:1985ag,Foot:1991kb}
incorporating a fermionic DM candidate provides
a natural extension of the SM of particle
physics to account for the dark matter sector \cite{Filippi:2020kii, Agrawal:2021dbo, Caputo:2021eaa}.
Under a dark $U(1)_D$ gauge symmetry \cite{Fabbrichesi:2020wbt}, the
coupling between the dark photon $A'_\mu$
and the DM particle $\chi$,
\begin{align}
  \mathcal{L}
\supset
  g_D A_\mu^\prime\bar{\chi}\gamma^\mu\chi,
\label{xxa}
\end{align}
takes a form similar to the electromagnetic interaction. Since the nature of the particle and its corresponding interactions have not yet been experimentally observed, there are almost no constraints on the dark gauge coupling $g_D$, whereas the kinetic mixing parameter is constrained \cite{Graham:2021ggy,Agrawal:2021dbo,Filippi:2020kii, Caputo:2021eaa,Lanfranchi:2020crw}. The sole constraint on $g_D$ arises from the Bullet Cluster and cosmological structure. We discuss this in detail in Sec.~\ref{sec:sidm}. Consequently, the dark gauge coupling $g_D$ can be large.

The CRDM may undergo further evolution via final-state radiation (FSR) if the DM energy and the characteristic energy scale of the hard process during acceleration are significantly larger than both the DM and mediator masses.

\subsection{Dark Splitting Functions}

With a sufficiently large dark gauge coupling, a dark parton shower inevitably occurs once a dark particle (either the dark photon or the dark fermion) is produced. This process can generate a chain of particles \cite{Chen:2018uii, Rizzo:2020jsm, Knapen:2021eip, Nam:2021bsf, Albouy:2022cin, Chigusa:2022act,Winkler:2022zdu,
 Li:2023fzv, Kulkarni:2024okx,
Kulkarni:2025rsl}. Such dark parton showers can yield rich phenomena in dark matter annihilation by promoting a suppressed $p$-wave process into a sizable $s$-wave one \cite{Bell:2017irk}, as well as facilitate possible detection through the dark trident channel \cite{deGouvea:2018cfv}. Furthermore, dark parton showers have been extensively explored at colliders \cite{Schwaller:2015gea, Cohen:2015toa,Bai:2015nfa, Buschmann:2015awa, Kim:2016fdv,Cohen:2020afv, Li:2021bka,Du:2021cmt,Bernreuther:2022jlj,Cohen:2023mya,Born:2023vll,Carrasco:2023loy,Cheng:2024hvq,Cheng:2024aco,Liu:2024rbe, Liu:2025bbc,Carmona:2024tkg,Li:2025tlg}.

In the presence of multiple external particles, calculating the amplitude and cross section using conventional Feynman diagram methods becomes highly challenging. One may instead employ parton shower techniques \cite{Collins:1989bt,Hoche:2014rga,Metz:2016swz,
Nagy:2017ggp, Papaefstathiou:2024qlg} to factorize the entire chain into a series of $1 \rightarrow 2$ splittings. The differential cross section for a hard process followed by an $A \to B + C$ branching can be factorized as
\begin{align}
  d \sigma_{X,BC}
\simeq
  d \sigma_{X,A}\times  d \mathcal{P}_{A \rightarrow B+C}~, 
\end{align}
where $X$ represents the additional particles in the final state of the hard process, excluding particle $A$.
The term $d \mathcal{P}_{A \rightarrow B+C}$ denotes the differential splitting function for the branching process $A \to B + C$,
\begin{align}
  \frac{d \mathcal{P}_{A \rightarrow B+C}}{d z ~d \ln Q^{2}}
\approx
  \frac{1}{N}
  \frac{1}{16 \pi^{2}}
  \frac{Q^2}{\left(Q^{2}-m_{A}^{2}\right)^{2}}
  \left|\mathcal M_{\text{split}}\right|^{2},
\label{split}
\end{align}
where $z$ is the energy fraction carried by $B$
and $Q^2$ is the virtuality carried by the intermediate particle $A$.
The squared matrix element ${\left|\mathcal M_{\text{split}}\right|^2}$ is calculated
using the amputated Feynman diagram for $A \to B + C$ with on-shell
polarization vectors.
The factor $N$ equals 2 for identical particles $B$ and $C$, and 1 otherwise.

In the context of a timelike branching process $A \rightarrow B + C$, we represent the particle momentum as
\begin{subequations}
\begin{align}
& P_A \equiv \left(E_A, 0, 0, E_A - \frac{k_T^2+\bar{z}m_B^2+zm_C^2}{2z\bar{z}E_A}\right), \\
&P_B \equiv \left(zE_A, k_T, 0, zE_A - \frac{k_T^2 + m_B^2}{2z E_A}\right), \\
&P_C \equiv \left(\bar{z}E_A, -k_T, 0, \bar{z}E_A - \frac{k_T^2 + m_C^2}{2\bar{z}E_A}\right),
\end{align}
\end{subequations}
where the energy fractions $z$ and $\bar{z} \equiv 1 - z$
lie within the interval $(0, 1)$. Assuming that $E_A^2$ is much larger than the transverse momentum squared $k_T^2$ and the mass squared $m_i^2$ for $i = A, B, C$, the corresponding virtualities are derived by neglecting terms of order $(k_T^2 \text{ or } m_i^2) / E_A^2$:
\begin{align}
 P_A^2 = Q^2=\frac{k_T^2+\bar{z}m_B^2+z m_C^2}{z\bar{z}},
\quad
 P_B^2 = m_B^2,
\quad
 P_C^2 = m_C^2.
\end{align}
While particles $B$ and $C$ satisfy the on-shell condition, particle $A$ exhibits a virtuality $Q$.

\begin{table}[htb]
	\centering
	\begin{tabular}{c|c}  
		$A\rightarrow B+C$&   $\frac{d \mathcal{P}_{A \rightarrow B+C}}{d z~ d\ln Q^{2}}=P_{A \rightarrow B+C}(z)$\\
		\hline 
        $A_{L}^{\prime} \rightarrow \bar{\chi} / \chi+\chi / \bar{\chi}$  & $\frac{2 \alpha^{\prime}}{\pi}  \frac{Q^2}{\left(Q^2-m_{A^\prime}^{2}\right)^{2}}m_{A^\prime}^{2} z \bar{z}$\\
		$A_{T}^{\prime} \rightarrow \bar{\chi} / \chi+\chi / \bar{\chi}$ & $\frac{\alpha^{\prime}}{2 \pi}\frac{Q^2}{\left(Q^{2}-m_{A^\prime}^{2}\right)^2}  \left(Q^2\left(z^{2}+\bar{z}^{2}\right)+2m_\chi^2\right)$\\
		$\chi / \bar{\chi} \rightarrow A_{L}^{\prime}+\chi / \bar{\chi}$ & $\frac{\alpha^{\prime}}{\pi}  \frac{Q^2 }{\left(Q^{2}-m_{\chi}^{2}\right)^{2}}m_{A^\prime}^{2} \frac{\bar{z}}{z^2}$\\
        $\chi / \bar{\chi} \rightarrow A_{T}^{\prime}+\chi / \bar{\chi}$ &   $\frac{\alpha^{\prime}}{2 \pi}\frac{Q^2}{\left(Q^{2}-m_{\chi}^{2}\right)^2}  \left(Q^2\frac{1+\bar{z}^{2}}{z}- m_{\chi}^{2}\frac{2+z^2}{z}-m_{A^\prime}^{2}\frac{1+\bar{z}^{2}}{z^2}\right)$\\
	\end{tabular}
	\caption{\label{tab:splitf} Splitting functions involving $A^\prime$ and $\chi/\bar{\chi}$.} 
\end{table}
The splitting functions for various timelike branching processes \cite{Metz:2016swz} are summarized in Table\,\ref{tab:splitf}. At a single vertex, the splitting function is proportional to the dark fine-structure constant $\alpha^\prime \equiv g_D^2 / 4\pi$. Because both the DM particle $\chi$ and the dark photon $A^\prime$ are massive, with masses $m_\chi$ and $m_{A'}$, respectively, the dark photon exhibits both transverse ($A^\prime_T$) and longitudinal ($A'_L$) polarization states. These splitting functions have been averaged over the polarizations of the initial particles and summed over the final states. When calculating splitting functions that include the longitudinal dark photon mode, it is crucial to omit terms proportional to $Q^2-m_A^2$ \cite{Chen:2016wkt}. The splitting function for the process $\chi / \bar{\chi} \rightarrow \chi / \bar{\chi} + A_{T/L}^{\prime}$ can be deduced from that for $\chi / \bar{\chi} \rightarrow A_{T/L}^{\prime} + \chi / \bar{\chi}$ via the relation $ P_{A\rightarrow B+C}(z) = P_{A\rightarrow C+B}(\bar{z})$.

\subsection{Final-State Parton Shower}
\label{subsec:FSR}

We compute the FSR evolution using a Monte Carlo method with a Markov chain governed by the Sudakov factors of DM $\chi$ and the mediator $A^\prime$ \cite{Hoche:2014rga,Metz:2016swz,Nagy:2017ggp,Papaefstathiou:2024qlg}. The evolution proceeds as follows:
\begin{enumerate}
\item \textbf{Initialization:} We begin at a high virtuality scale $Q_{\text{max}}$, which we set equal to the momentum transfer $\sqrt{2 m_\chi T_\chi}$, where $T_\chi$ denotes the kinetic energy of the boosted DM in the hard DM–cosmic ray scattering process. In our simulation, we require the DM particle to possess a kinetic energy $T_\chi > T_{\chi,\text{min}}^{\text{FSR}}\equiv (m_\chi + m_{A'})^2 / 2 m_\chi$ prior to the FSR stage. This requirement follows from the condition $Q_{\text{max}} > m_\chi+m_{A'}$, which ensures that the $\chi \rightarrow A' + \chi$ splitting is kinematically allowed immediately after the hard DM–cosmic ray scattering.

\item \textbf{Sudakov Factor:}
The logarithmic evolution step is employed to refine the simulation in the low-virtuality region. 
Within the probabilistic framework of the parton shower, the Sudakov form factor,
\begin{align}
  \Delta_A(Q_2;Q_1)
\equiv
  \exp\left[-\sum_{BC}\int_{\ln Q_1^2}^{\ln Q_2^2}d\ln Q^2 \int_{z_{\text{min}}(Q)}^{z_{\text{max}}(Q)} dz ~\frac{d \mathcal{P}_{A \rightarrow B+C}(z,Q)}{dz~d\ln Q^2}\right],
\end{align}
plays a pivotal role.
This factor determines the probability that a parton
$A$ does not undergo branching as the virtuality scale
$Q$ evolves from $Q_2$ to $Q_1$, where $Q_2 > Q_1$.
We also define a low-virtuality cutoff $Q_{\text{min}}$,
below which the parton shower evolution terminates. This cutoff is typically chosen to be on the order of the dark particle masses: $Q_\text{min}\equiv m_\chi+m_{A'}$ for the $\chi \rightarrow A + \chi$ splitting, or $2 m_\chi$ for the $A \rightarrow \chi + \bar \chi$ splitting. 

A random number $R$ is drawn from a uniform distribution between 0 and 1. The branching probability, $P_{\text{Branch}}$, is calculated from the Sudakov factor and the relevant splitting function, integrated over the appropriate phase space. If $R < P_{\text{Branch}}$, the branching $A \to B + C$ occurs; otherwise, the parton continues to evolve to a lower virtuality scale without branching at this step.
This procedure applies to the determination of both $Q^2$
in the current step and the energy fraction $z$ in the subsequent step.

\item \textbf{Branching Kinematics:} At each step in the evolution of $Q$, we compute the probability that parton $A$ branches into daughter partons $B$ and $C$ ($A \to B + C$). The kinematically allowed range for $z$, denoted by $(z_{\text{min}}(Q), z_{\text{max}}(Q))$ at a given scale $Q$,
\begin{subequations}
\begin{align}
& z_{\text{min}}(Q)
\equiv
  \frac{Q^2 + m_B^2 - m_C^2 - \sqrt{(Q^2 - m_B^2 - m_C^2)^2 - 4m_B^2 m_C^2}}{2Q^2},
\\
& z_{\text{max}}(Q)
\equiv
  \frac{Q^2 + m_B^2 - m_C^2 + \sqrt{(Q^2 - m_B^2 - m_C^2)^2 - 4m_B^2 m_C^2}}{2Q^2},
\end{align}
\end{subequations}
is influenced by the kinematic conditions.
This probability is given by the splitting function in Eq.\,(\ref{split}), with model-specific details provided in Table~\ref{tab:splitf}.

\item \textbf{Recursive Evolution:} 
If a branching $A \to B + C$ occurs at a virtuality scale $Q$, the parton shower evolution continues recursively and independently for both daughter partons, $B$ and $C$. Each daughter parton is treated as a new parent, and the evolution process is repeated starting from the scale $Q$. Angular ordering is implemented by imposing a veto on subsequent branchings: a splitting is rejected if the opening angle between the daughter partons is greater than that of the parent splitting.

\item \textbf{Kinematic rearrangement:} 
During the splitting $A \to B + C$, particle $A$ acquires virtuality, violating energy and momentum conservation. To address this, a dipole recoil scheme \cite{Gustafson:1987rq, Sjostrand:2006za} is employed. In this scheme, $X A$ is treated as the initial dipole, and the energies and momenta of both $X$ and $A$ are reset in their center-of-mass frame while preserving the center-of-mass energy. Following this kinematic rearrangement, a boost transforms the momenta of $X$, $B$, and $C$ back to the original laboratory frame. 

\end{enumerate}

\subsection{FSR Evolution Kernel}
\label{subsec:kernel}

\begin{figure}[htb]
    \centering
        \includegraphics[width=0.47\textwidth]
        {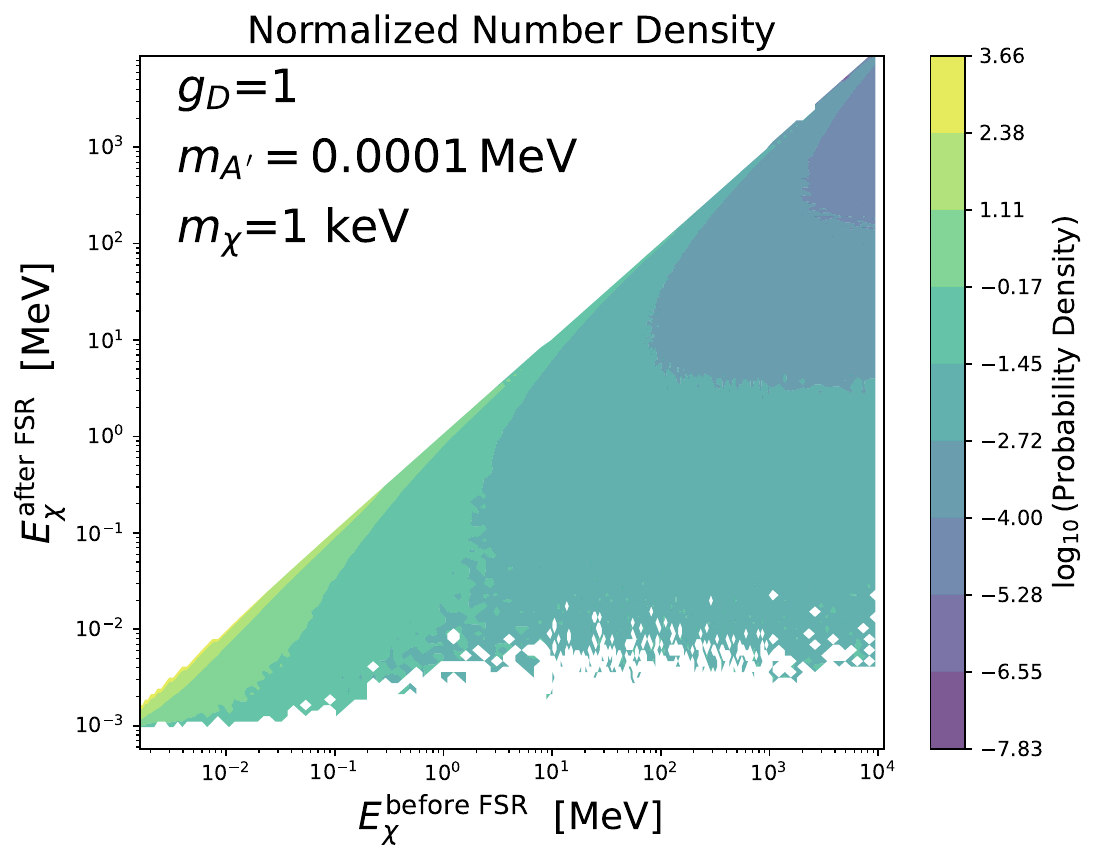}
        \includegraphics[width=0.47\textwidth]
        {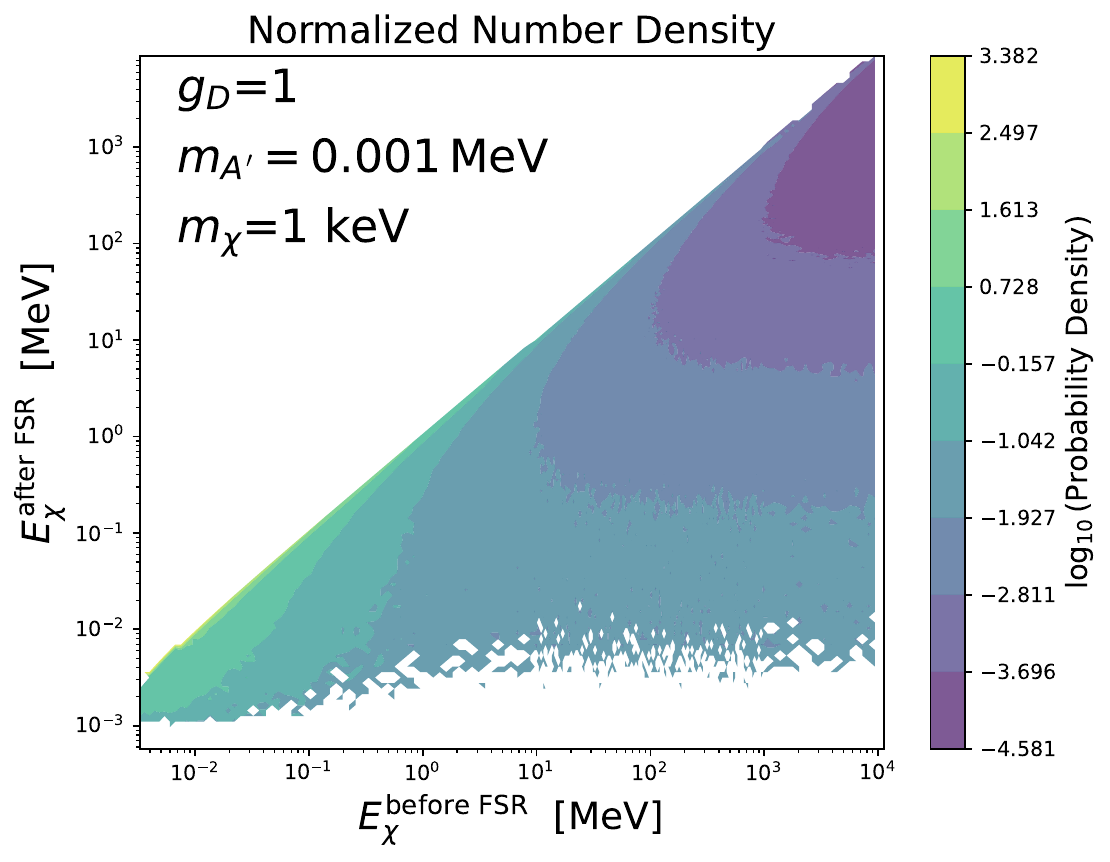}\\
        \includegraphics[width=0.47\textwidth]
        {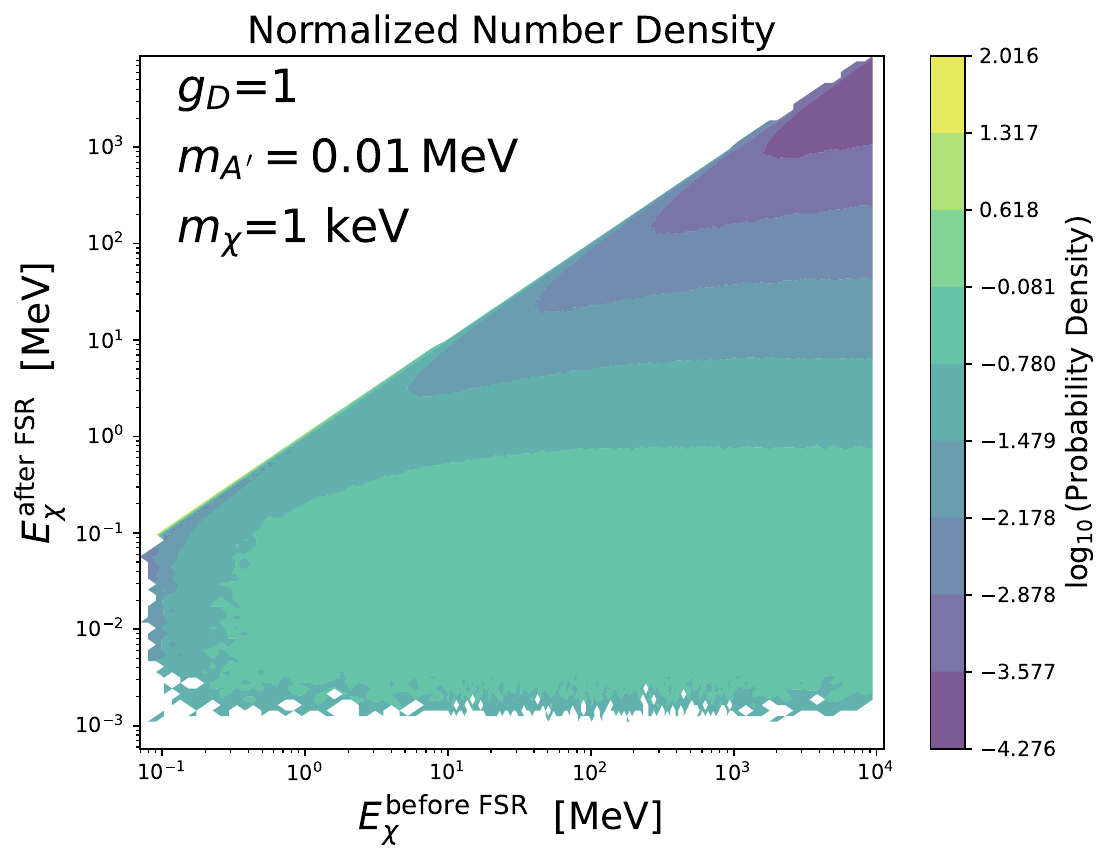}
        \includegraphics[width=0.47\textwidth]
        {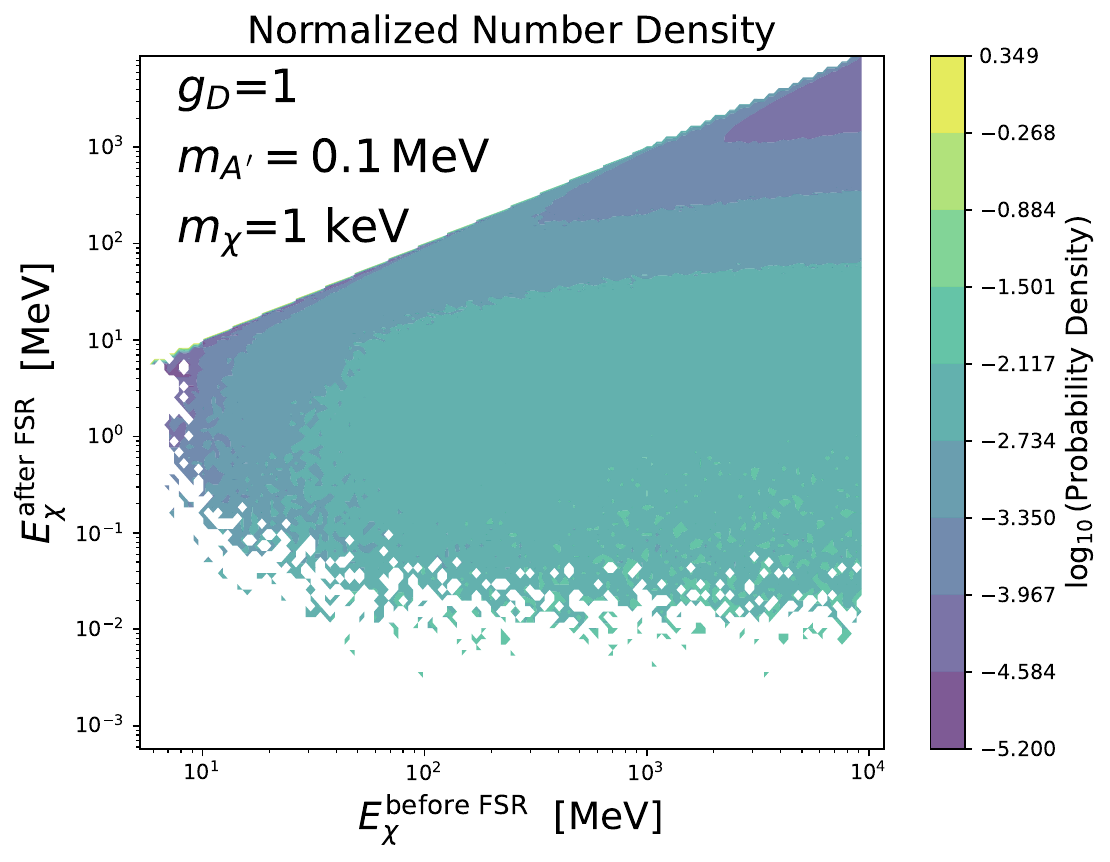}
    \caption{\label{fsrkernel1}The normalized number density of DM particles after FSR for $g_D=1$. The DM mass is fixed at 1\,keV, while the mediator masses are indicated in the corresponding plots.}
\end{figure}

\begin{figure}[htb]
    \centering
        \includegraphics[width=0.47\textwidth]
        {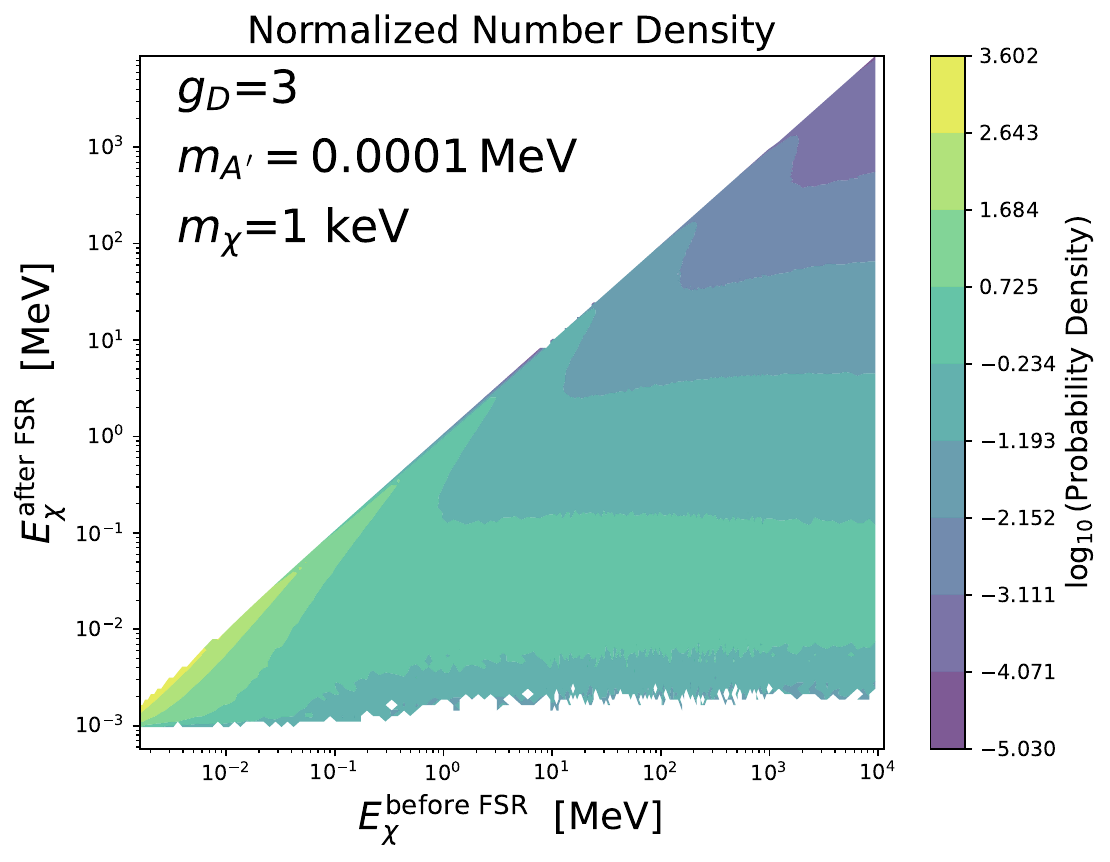}
        \includegraphics[width=0.47\textwidth]
        {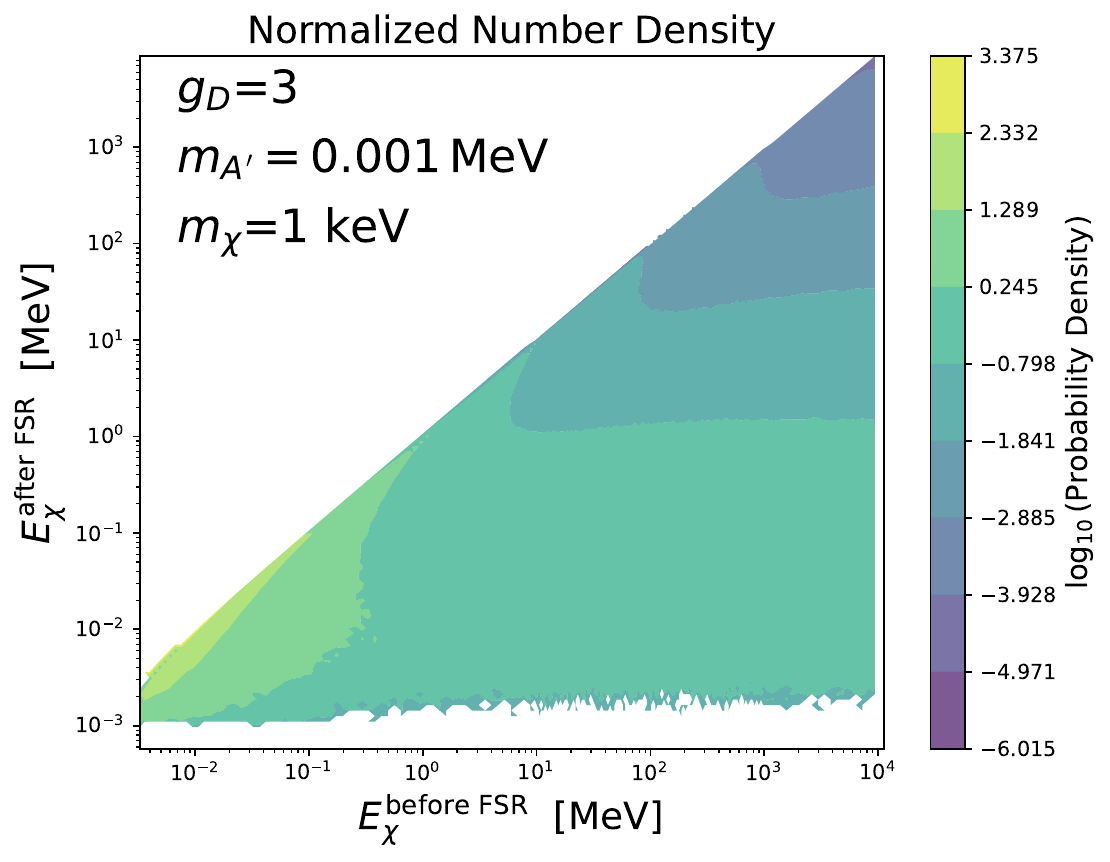}\\
        \includegraphics[width=0.47\textwidth]
        {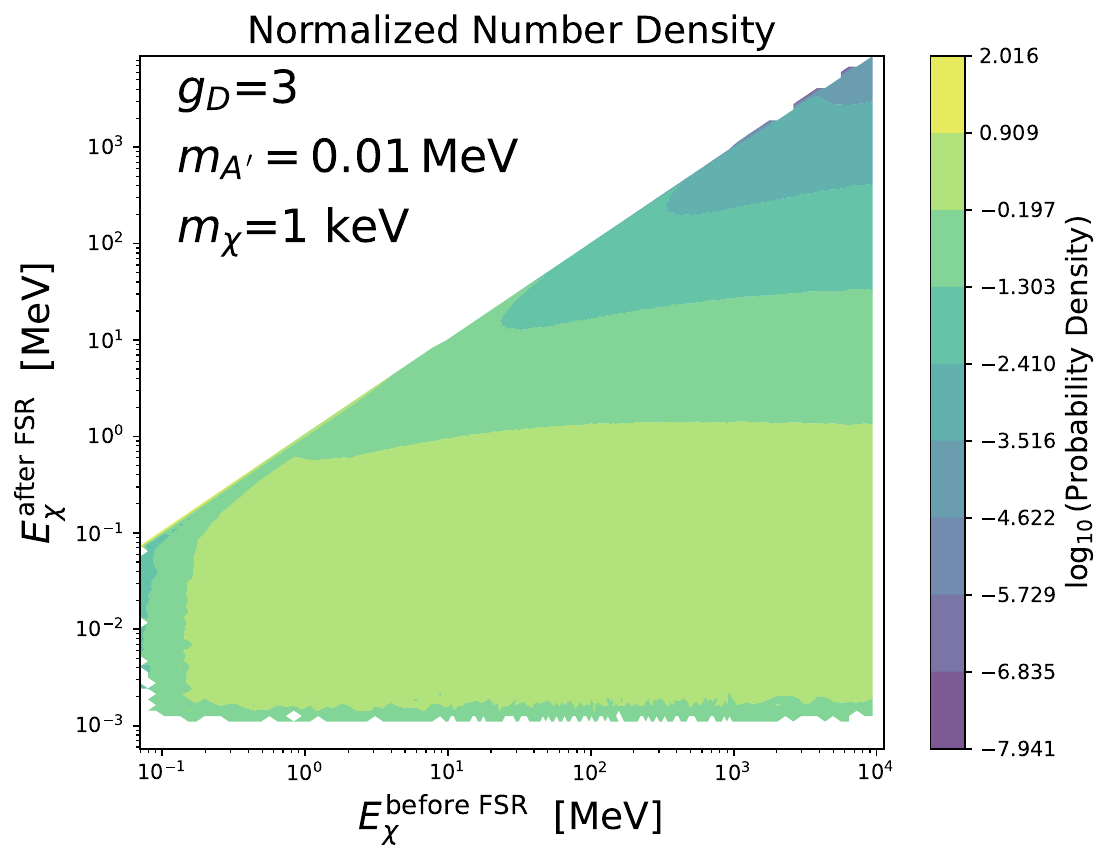}
        \includegraphics[width=0.47\textwidth]
        {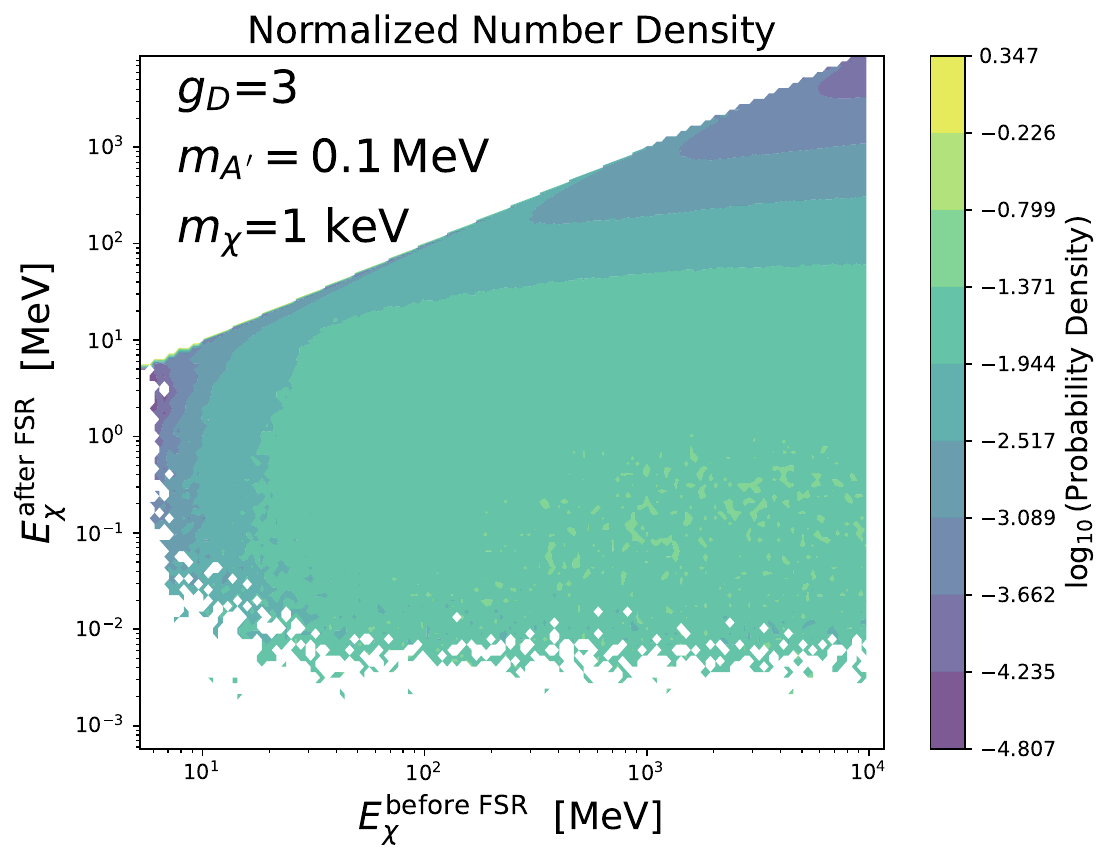}
    \caption{\label{fsrkernel3}The normalized number density of DM particles after FSR for $g_D=3$. The DM mass is fixed at 1\,keV, while the mediator masses are indicated in the respective plots.}
\end{figure}

Using the numerical simulations described above, we determine the evolution kernel $\mathcal{F}(E^0_\chi ,E_\chi)$ for the FSR process. This kernel characterizes the probability distribution for the energy $E_\chi$ of the final-state $\chi$ particles produced by the splitting of an initial CRDM particle with energy $E^0_\chi$. 
The average number of $\chi$ particles produced following FSR from an initial $\chi$ with kinetic energy $E^0_\chi$ is given by
\begin{align} \label{eq:kernel}
N_\chi^{\text{FSR}} (E_\chi^0) = \int \mathcal{F}(E^0_\chi ,E_\chi) dE_\chi~.~
\end{align}
Furthermore, the final kinetic energy spectrum of the CRDM particles, including the effects of FSR, can be obtained by convolving the initial flux with the evolution kernel,
\begin{align}
  \frac{d \Phi_\chi}{d T_\chi}
=
  \int \frac{d \Phi^0_{\chi}}{d T^0_\chi}
  \mathcal{F}
\left(
  {T^0_\chi+m_\chi},
  {T_\chi+m_\chi}
\right) d T_\chi^0~,~
\label{fluxfsr}
\end{align}
where $d\Phi^0_\chi / dT^0_\chi$ represents the initial kinetic energy spectrum of the CRDM before FSR, as calculated in Eq.\,(\ref{flux}). Note that $d \Phi_\chi / d T_\chi$ and $\mathcal{F}(E^0_\chi ,E_\chi)$ also include the anti-DM component. This is because anti-DM particles are produced via FSR.  

The FSR evolution kernels $\mathcal{F}(E_\chi^{\text{before FSR}}, E_\chi^{\text{after FSR}})$
are plotted in Figures~\ref{fsrkernel1} and \ref{fsrkernel3} for DM couplings $g_D=1$ and 3, respectively.  
As defined in Eq.\,(\ref{eq:kernel}), the evolution kernel has units of MeV$^{-1}$.
We take the DM mass $m_\chi = 1$\,keV hereafter as a representative value. As the dark photon mass increases,  the minimum value  of  $E_\chi^{\text{before FSR}}$ required for FSR to occur also increases, as  discussed above.\footnote{Here, we use the total energy $E_\chi$ instead of the kinetic energy $T_\chi$. The minimum value of $E_\chi^{\text{before FSR}}$ is $T_{\chi,\text{min}}^{\text{FSR}} +m_\chi$.}  For instance, a CRDM particle with an energy of 10\,keV can undergo FSR only if $m_{A'} \lesssim 3.2$\,keV, whereas the threshold energy rises to approximately 5\,MeV for $m_{A^\prime} = 0.1$\,MeV.   The FSR effects are more significant for larger $g_D$, as seen by comparing the results in Figures~\ref{fsrkernel1} and \ref{fsrkernel3}. Additionally, some plots exhibit blank regions at keV-scale $E_\chi^{\text{after FSR}}$ values. This occurs because the integrated number density over $E_\chi^{\text{after FSR}}$ in this range is too small, resulting in an absence of FSR events in the Monte Carlo simulation.

For each fixed $E_\chi^{\text{before FSR}}$ in the plots,  the vertical profile displays the number density as a function of
 $E_\chi^{\text{after FSR}}$.  As  $E_\chi^{\text{before FSR}}$ increases,  the density at the upper edge, corresponding to  $E_\chi^{\text{after FSR}}=E_\chi^{\text{before FSR}}$, decreases, and the peak of the distribution gradually shifts to lower $E_\chi^{\text{after FSR}}$.  This behavior indicates that FSR effects become more significant for higher-energy CRDM particles. Moreover, the FSR contribution is suppressed as the dark photon mass $m_{A'}$ increases. This is evident when comparing the results for $m_{A'}=0.1$\,MeV to those with smaller $m_{A'}$: the $E_\chi^{\text{before FSR}}$ threshold for FSR becomes significantly higher, and the overall number density after FSR is noticeably reduced.  

Additionally, we highlight a subtle feature that is not immediately visible in the plots.  This feature occurs in the parameter space where $m_{A'}<10^{-3}$\,MeV, $E_\chi^{\text{before FSR}}$ is large ($>\mathcal{O}(10)$\,MeV), and $E_\chi^{\text{after FSR}}
\approx E_\chi^{\text{before FSR}}$ (i.e., near the edge of the plots).  In this region,  for a fixed $E_\chi^{\text{before FSR}}$, the number density  decreases as $m_{A'}$ increases.  This can be explained by the fact that a relatively heavier dark photon carries away energy from the initial CRDM particle more efficiently. This feature is important for understanding why FSR leads to more significant modifications of the exclusion bounds for keV-scale dark photons in detectors like Super-Kamiokande, where the relevant energy scale is  $\mathcal{O}(100)$\,MeV, as will be discussed in Section 
\ref{subsec:bound}.

\section{Boosted Dark Matter from Cosmic Ray Acceleration}
\label{sec3}

In our simplified model, the dark photon $A^\prime$ kinetically mixes with the SM photon. This kinetic mixing induces an effective coupling between the dark photon and SM fermions, given by~\cite{Holdom:1985ag,Foot:1991kb}:
\begin{align}
\mathcal{L} \supset \epsilon g_{\text{em}} A^\prime_\mu \bar{e}\gamma^\mu e ~,~ \label{eq:model2}
\end{align}
where $\epsilon$ parameterizes the kinetic mixing strength. For illustrative purposes, we assume that the dark photon couples predominantly to electrons. This scenario is widely adopted in interpretations of the PAMELA and DAMPE data~\cite{PAMELA:2008gwm,DAMPE:2017fbg,Fox:2008kb,Bi:2009uj,Gu:2017gle,Duan:2017pkq,Chao:2017emq,Duan:2017qwj,Ghorbani:2017cey,Ge:2017tkd}, which reported excesses in the electron-positron cosmic-ray spectrum. However, incorporating couplings to other SM fermions would increase the boosted dark matter flux, thereby enhancing the dark parton shower signature central to this work. Our setup is therefore conservative. The most stringent exclusion limits on the kinetic mixing parameter $\epsilon$ arise from stellar cooling~\cite{Antel:2023hkf} and beam-dump experiments that employ the ``missing energy'' technique to probe the invisible decay of the $A^\prime$~\cite{NA64:2023wbi, Banerjee:2019pds, Ge:2025aui}. We address these constraints on the relevant parameter space in the conclusion (see Refs.~\cite{Filippi:2020kii, Agrawal:2021dbo, Caputo:2021eaa, Fabbrichesi:2020wbt} for reviews).

With the dark photon $A'$ mediating interactions between the DM particle $\chi$ and SM particles, non-relativistic halo DM particles can be naturally accelerated by energetic cosmic rays in the Milky Way~\cite{Yin:2018yjn,Cappiello:2018hsu,Bringmann:2018cvk,Ema:2018bih}. Under the assumptions of a homogeneous CR distribution and an NFW DM halo profile~\cite{Navarro:1995iw,Navarro:1996gj} with $\rho^\mathrm{local}_\chi \sim 0.4~\mathrm{GeV}~\mathrm{cm}^{-3}$~\cite{Salucci:2010qr,Fermi-LAT:2012pls}, the differential recoil flux of CRDM is given by~\cite{Li:2022dqa, Bondarenko:2019vrb}
\begin{equation}\label{flux}
\frac{d \Phi_{\chi}^0}{d T_{\chi}}=D_{\mathrm{eff}} \frac{\rho_{\chi}^{\text {local }}}{m_{\chi}} \int_{T_{\mathrm{CR}}^{\min }}^{\infty} d T_{\mathrm{CR}} \frac{d \Phi_{e}}{d T_{\mathrm{CR}}} \frac{d \sigma_{\chi e}}{d T_{\chi}}.
\end{equation}
The CR flux $d\Phi_e / dT_\mathrm{CR}$ is simulated using HelMod-4 \cite{Boschini:2018zdv}. The effective distance $D_\mathrm{eff}=8.02$\,kpc is determined by integrating along the line of sight up to 10\,kpc \cite{Bringmann:2018cvk}. In the 2-to-2 scattering process, the initial halo DM is assumed to be at rest, which is a good approximation given the large momentum transfer. Therefore, the differential cross section can be expressed as \cite{Li:2022dqa, Cao:2020bwd},
\begin{equation}
\label{cr-crossing}
  \frac{d \sigma_{\chi e}}{d T_{\chi}}
=
  g_D^{2} (\epsilon g_{\rm em})^{2}
  \frac {2 m_{\chi} \left(m_{e} + T_{\mathrm{CR}}\right)^{2}
        -T_{\chi} \left[ \left(m_{e}+m_{\chi}\right)^{2}+2 m_{\chi} T_{\mathrm{CR}} \right]
        + m_{\chi} T_{\chi}^{2}}
        {4 \pi\left(2 m_{e} T_{\mathrm{CR}}+T_{\mathrm{CR}}^{2}\right)\left(2 m_{\chi} T_{\chi}+m_{A}^{2}\right)^{2}}~.~
\end{equation}
Moreover, the minimum incoming kinetic energy of a cosmic electron in Eq.~(\ref{flux}) is given in Ref.~\cite{Li:2022dqa, Cao:2020bwd}.
\begin{equation}
  T_{\mathrm{CR}}^{\min }
=
  \left( \frac{T_{\chi}} 2 - m_e \right)
\left[
  1
\pm
  \sqrt{1+\frac{2 T_{\chi}}{m_{\chi}} \frac{\left(m_{e}+m_{\chi}\right)^{2}}{\left(2 m_{e}-T_{\chi}\right)^{2}}}
\right],
\end{equation} 
where the $+$ sign corresponds to $T_\chi > 2 m_e$ and the $-$ sign to $T_\chi < 2 m_e$.

\subsection{Dark matter flux before and after the splitting}\label{subsec:flux}

The FSR effects can be incorporated into the CRDM flux calculation using Eqs.\,(\ref{flux}) and (\ref{fluxfsr}), where the FSR evolution kernel is derived from Monte Carlo simulations.
\begin{figure}[tb]
\centering
\includegraphics[width=0.45\textwidth]{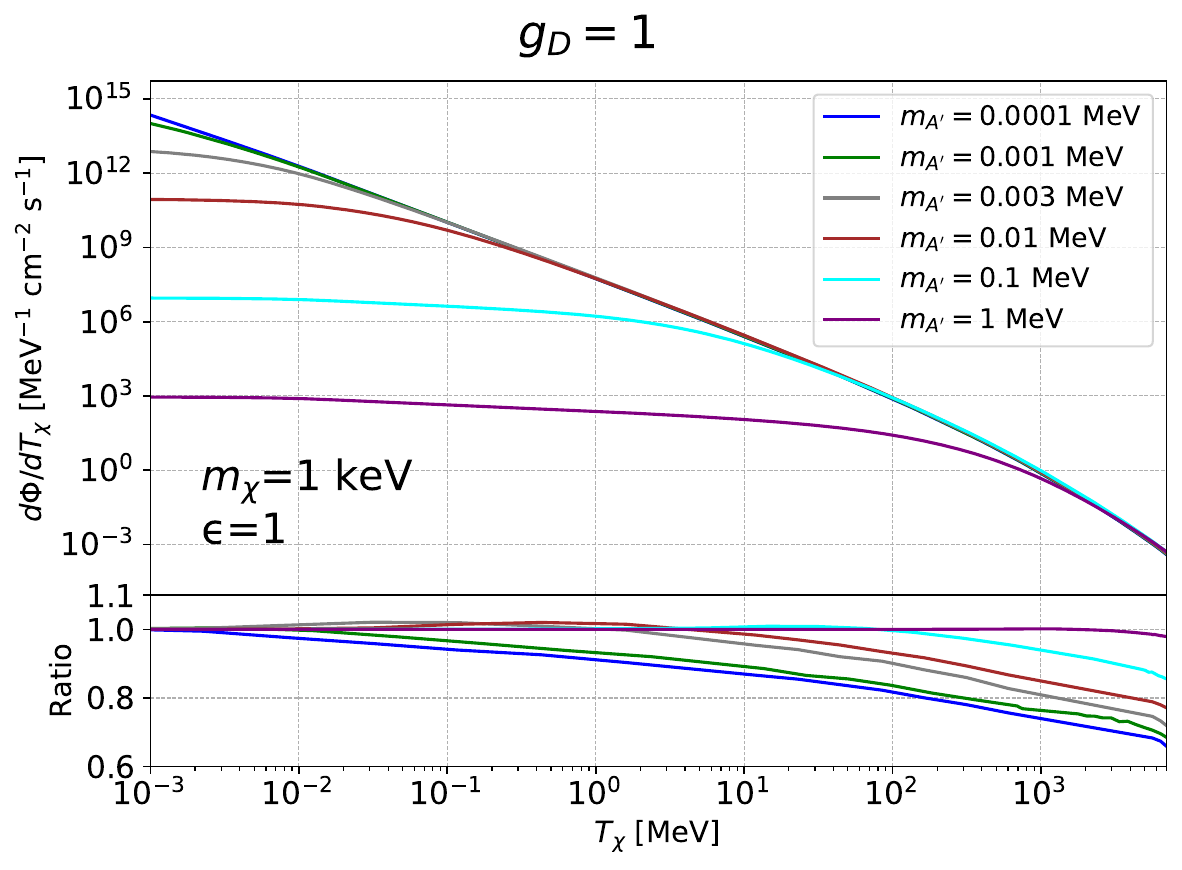}
\includegraphics[width=0.45\textwidth]{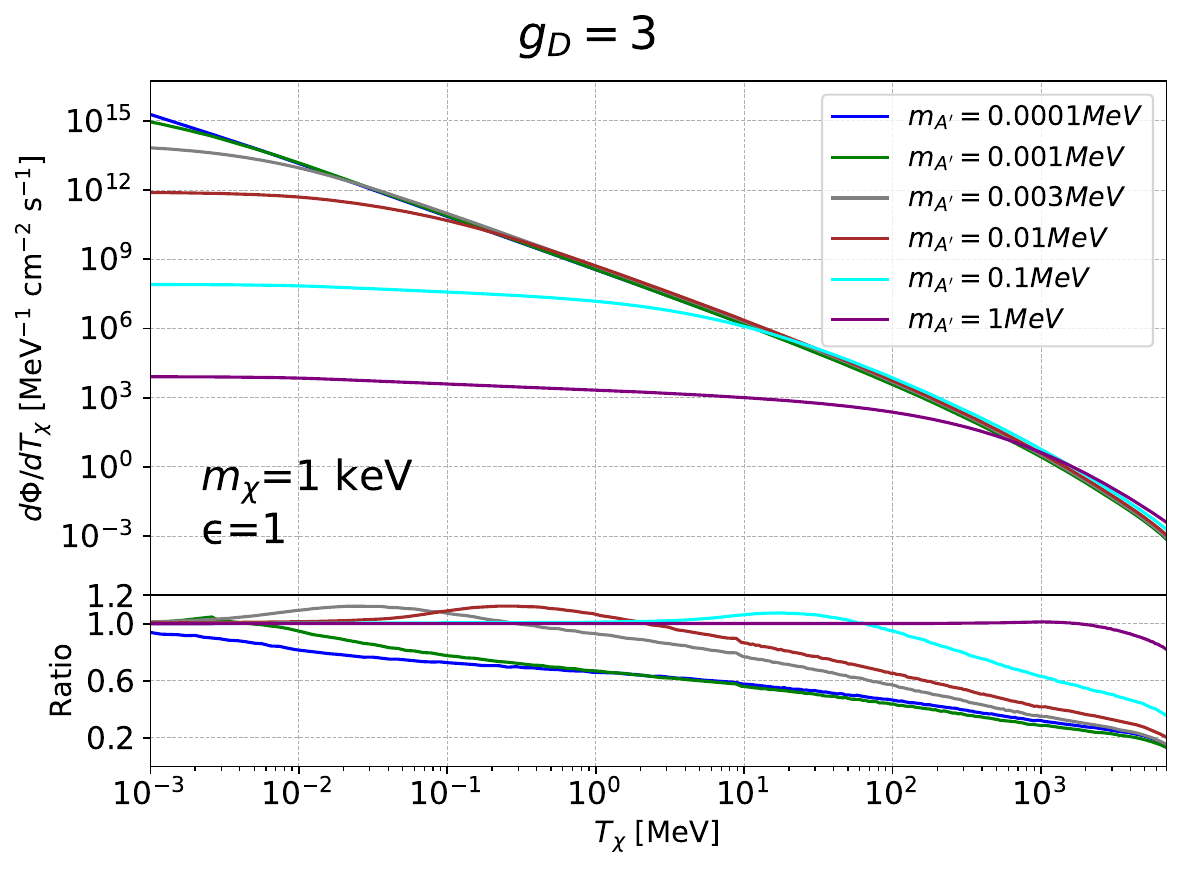}
\caption{The CRDM fluxes including FSR effects (upper panels) and the ratios of the fluxes with FSR to those without FSR (lower panels) are presented for $g_D = 1$ (left) and $g_D = 3$ (right), respectively. The dark photon mass values are indicated in each panel. For illustrative purposes, we fix $m_\chi=1$\,keV and $\epsilon=1$.
}
\label{fig:fluxl1}
\end{figure}

The upper panels in Figure~\ref{fig:fluxl1} show the differential CRDM fluxes including FSR effects for $g_D=1$ (left) and $g_D=3$ (right). The lower panels display the ratio between the DM fluxes with and without FSR. The overall flux scales proportionally to $\epsilon^2$; thus, fluxes for other values of $\epsilon$ can be obtained via simple rescaling. Although the pre-FSR flux (the flux without FSR) is not explicitly shown, its value can be inferred from the ratios in the lower panels. In the small $T_\chi$ region, the pre-FSR flux is suppressed as the dark photon mass $m_{A'}$ increases, whereas in the large $T_\chi$ region, it is largely independent of $m_{A'}$. Consequently, the DM flux shape flattens as $m_{A'}$ increases. This feature is crucial for understanding the impact of FSR.

FSR predominantly depletes the high-energy CRDM flux through dark photon emission. When kinematically allowed, the subsequent splitting or decay of these dark photons produces a significant number of secondary DM particles with lower kinetic energy. Overall, FSR modifies the flux at a given $T_\chi$ in two competing ways: it reduces the energy of DM particles originally at $T_\chi$, thereby decreasing the flux at that energy, while simultaneously producing additional DM particles at $T_\chi$ from the FSR of higher-energy DM. The net effect at $T_\chi$ depends strongly on the shape of the pre-FSR flux. For the energy range of interest, $T_\chi \sim \mathcal O($\,keV$-$MeV$)$, a flatter pre-FSR flux generally yields a net enhancement after FSR, whereas a steeply falling pre-FSR flux results in a net suppression.

In the lower panels of Figure\,\ref{fig:fluxl1}, the FSR effects are similar for both $g_D=1$ and $g_D=3$, but the features are more pronounced for the larger $g_D$ value. For larger mediator masses $m_{A'}$, where the decay $A' \to \chi \chi$ is kinematically allowed, FSR yields a slight enhancement of the flux over a specific $T_\chi$ range, producing a localized bump. Below this range, the flux remains largely unchanged, while at higher energies it is suppressed. Conversely, for smaller values of $m_{A'}$, the outcome is a net reduction of the flux. These features, observed for both large and small $m_{A'}$, arise from three primary factors: the shape of the pre-FSR dark matter flux, the expansion of the showering phase space, and the kinematic accessibility of the $A' \to \chi \chi$ decay.

To illustrate the impact of these primary factors on FSR, we examine two benchmark points. The first is characterized by $m_{A'} = 0.1$\,MeV and $g_D=3$, where the decay channel is open and the kinetic energy threshold for FSR is $T_{\chi,\text{min}}^{\text{FSR}} \sim 5$ MeV, as explained in Sec.~\ref{subsec:FSR}. The post-FSR flux at $T_\chi < 0.1$ MeV remains almost unchanged relative to the pre-FSR flux. This occurs because, in this region, FSR affects the flux only through the emission of secondary DM particles from ancestor DM with $T_\chi>5$ MeV, whose flux is at least 11 times smaller than that at $T_\chi < 0.1$ MeV. The post-FSR flux at $T_\chi > 65$ MeV is reduced compared to the pre-FSR flux due to the steeply falling spectrum in this regime, where the depletion of DM from FSR exceeds the contribution from secondary DM produced by higher-energy DM particles. Conversely, an enhancement by a factor of up to 1.08 is observed only within the intermediate energy range of $T_\chi \sim \mathcal{O}(1)-\mathcal{O}(10)$\,MeV, owing to the relatively flat pre-FSR flux in this region. The second benchmark considers $m_{A'} = 10^{-4}$\,MeV and $g_D=3$, which lowers the threshold to $T_{\chi,\text{min}}^{\text{FSR}} \sim 0.6$\,keV. Here, the pre-FSR spectrum already falls rapidly throughout the keV scale. Since DM particles with kinetic energies at this scale are now kinematically permitted to shower, FSR induces a further reduction of the flux, even for $T_\chi$ at the $\mathcal{O}(1)$\,keV scale.

Furthermore, examination of the lower right panel for $g_D=3$ reveals a subtle detail in the high-energy region ($T_\chi>10$\,MeV). Comparing the results for $m_{A'}=10^{-4}$\,MeV (blue line), $m_{A'}=10^{-3}$\,MeV (green line), and $m_{A'}=3\times 10^{-3}$\,MeV (grey line), we observe that the FSR effects are slightly more pronounced for $m_{A'}=10^{-3}$\,MeV than in the other two scenarios. This occurs although their pre-FSR fluxes in this high-energy region are nearly identical, and their respective $T_{\chi,\text{min}}^{\text{FSR}}$ values are negligible compared to 10 MeV. This phenomenon can be attributed to two countervailing factors. On the one hand, as explained in Section\,\ref{subsec:kernel}, the emission of a more massive dark photon (1 keV) more effectively softens the energy of CRDM particles compared to the emission of a less massive one (0.1 keV). On the other hand, at $m_{A'}=3$\,keV, the decay channel for a dark photon to decay into a dark matter pair becomes kinematically accessible, which mitigates the overall suppression of the flux.

\section{Boosted Dark Matter Scattering with Electron}
\label{sec4}

\subsection{Dark matter direct detection experiments}

As discussed in recent studies~\cite{Bringmann:2018cvk,Wang:2019jtk,Ge:2020yuf,Cao:2020bwd,Jho:2020sku,Lei:2020mii,Flambaum:2020xxo,Dent:2019krz,Cho:2020mnc,Xia:2020apm,Feng:2021hyz,Wang:2021nbf,Emken:2021vmf,Bardhan:2022bdg,Alvey:2022pad,Xu:2024iny}, direct detection experiments targeting CRDM provide new access to the low-mass parameter space, whereas traditional methods focusing on non-relativistic halo DM via keV electron recoils lack sensitivity to sub-GeV dark matter. The complete ionization process, in which DM scatters off a target atom ($A$), is described by $\chi+A \rightarrow \chi+A^{+}+e^{-}$. This process can be simplified to $\chi(p_1)+e^-(p_2) \rightarrow \chi(k_1)+e^{-}(k_2)$ by treating the initial electron as a bound state and the final electron as free. This simplified scattering framework has been studied in Ref.\,\cite{Li:2022dqa}. We adopt the same method and parameterization for the kinematic phase space. Therefore, the differential cross section with respect to the electron recoil kinetic energy $T_R$ is given by
\begin{align}
\frac{d \sigma_{nl}}{d\ln T_R} =\frac{2l+1}{16 \cdot (2\pi)^5} & \frac{T_R |\bf{p_2}|}{E_\chi(m_e-E_B^{nl}) |{\bf p_1}| } \left|i \mathcal M\left({p_{1}}, {p_{2}},{k_{1}}, {k_{2}}\right)\right|^{2} \nonumber \\
&\times  |\chi_{n l }(|{\bf p_2}|)|^2d\phi_{p_2} d| {\bf p_2}| dq ~,~ \label{sigma1}
\end{align}
where $E_B^{nl}$ and $E_\chi$ represent the binding energy of the $(n,l)$ electron shell of the atom and the initial DM energy, respectively. The radial wave function $\chi_{nl}(|{\bf p_2}|)$ in momentum space for an electron in a xenon atom is taken from Ref.~\cite{Bunge:1993jsz}. The amplitude $\left|i  \mathcal M \left({p_{1}}, {p_{2}},{k_{1}}, {k_{2}}\right)\right|$ describes the scattering of DM off a bound electron, with an effective mass~\cite{Whittingham:2021mdw} given by $m_\mathrm{eff}^2 \equiv (m_e-E_B^{nl})^2-|{\bf p_2}|^2$.
Additionally, $q$ denotes the momentum transfer during the scattering process. The explicit expression for this amplitude, along with the integration ranges, can be found in Ref.\,\cite{Li:2022dqa}.

With the post-FSR CRDM flux and the differential scattering cross section determined, the resulting differential ionization rate\footnote{Here, we assume that the light dark photon ($m_{A'}<2m_\chi$) decays swiftly into SM particles with a lifetime too short to reach Earth; thus, the contribution to the ionization rate from dark Compton scattering ($A'~e \rightarrow \gamma ~e$) is not included.} is given by
\begin{equation}\label{eq:ddrecoil}
\begin{split}
\frac{dR_{ion}}{d\ln T_R}&=\sum_{nl}N_T\int dT_\chi\frac{d\sigma_{nl}}{d\ln T_R}\frac{d\Phi_\chi}{dT_\chi}~,~
\end{split}
\end{equation} 
where $N_T$ denotes the total number of target atoms.

\subsection{Neutrino detector - higher threshold}

Super-K~\cite{Super-Kamiokande:2002weg} is a water Cherenkov detector capable of probing recoil electrons with kinetic energies exceeding 100\,MeV. Because of the large momentum transfer during scattering, the initial-state atomic electron can be approximated as a free particle at rest. Consequently, the cross section for the process $\chi(p_1)+e^-(p_2)\rightarrow \chi(k_1)+e^-(k_2)$ becomes
\begin{equation}\label{sk-crossing}
\begin{split}
\frac{d\sigma}{d\ln T_R}&=\frac{1}{32\pi}\frac{T_R}{\left|\bf{p_1}\right|E_\chi m_e }|i \mathcal M_{\chi e}|^2,
\end{split}
\end{equation}
where the amplitude, expressed as a function of the Mandelstam variables, is given by
\begin{align}\label{sk-amp}
  |i \mathcal M_{\chi e}|^{2}
\equiv
  2 g_D^{ 2}(\epsilon g_{\text{em}})^2\frac{2(s(t-2m_e^2-2m_\chi^2)+(m_e^2+m_\chi^2)^2+s^2)+t^2}{(t-m_{A^{\prime}}^2)^2}.
\end{align}

The ionization rate is identical to that given by Eq.\,(\ref{eq:ddrecoil}), but without differentiating between electrons in different shells:
\begin{align}\label{eq:vdrecoil}
\frac{dR_{ion}}{d\ln T_R}&= N_{e}\int dT_\chi\frac{d\sigma}{d\ln T_R}\frac{d\Phi_\chi}{dT_\chi}.
\end{align}
 The lower limit of the initial DM kinetic energy, $T_\chi$, in Eq.~(\ref{eq:vdrecoil}) is
 \begin{equation}\label{range-sk-1}
\begin{split}
T_\chi>T_\chi^{\min}& \equiv \left(\frac{T_{R}}{2}-m_{\chi}\right)\left[1 \pm \sqrt{1+\frac{2 T_{R}}{m_{e}} \frac{\left(m_{e}+m_{\chi}\right)^{2}}{\left(2 m_{\chi}-T_{R}\right)^{2}}}\right]~,~
\end{split}
\end{equation}
where the $+(-)$ sign corresponds to $T_R>2m_\chi$ ($T_R<2m_\chi$). The explicit derivation of Eqs.\,(\ref{sk-crossing}), (\ref{sk-amp}), and (\ref{range-sk-1}) can be found in Ref.\,\cite{Li:2022dqa}.    

Because the hierarchy between the DM/dark photon mass and the DM-electron scattering energy scale also remains significant at neutrino detectors, dark matter PDFs must be considered when calculating ionization rates, as demonstrated in our previous study~\cite{Li:2022dqa}.  
Under the interaction in Eq.~\ref{xxa}, boosted DM can also generate dark photon and anti-DM densities, which are determined by solving the DGLAP equation. 
All these components contribute to DM-electron scattering. 
Following the formalism in Ref.~\cite{Li:2022dqa}, the differential cross section is modified by the PDFs as follows:
\begin{equation}
\frac{d \sigma}{d \ln T_{R}}=\sum_{i} \int_{0}^{x_{\max}} d x \frac{d \sigma^{i}}{d \ln T_{R}} f_{i}\left(Q, x\right) \Theta(xE^0_\chi-E_i^{\min}) ~.~
\end{equation}
The index $i$ runs over DM ($\chi$), anti-DM ($\bar{\chi}$), and the dark photon ($A^{\prime}$). The energy fraction $x$ is determined by the incident DM energy $E_\chi^0$ and the masses of the dark sector particles, whereas $E_i^\text{min}$ is governed by the kinematics of the hard scattering process. As explained in Ref.~\cite{Li:2022dqa}, an additional correction must be included, in which the integrated parton density $\int_{x_\text{max}}^1 f_{i}\left(Q, x\right) dx$ is interpreted as the probability of the DM having $x=1$ (i.e., the DM carrying all the initial energy). Further details are provided in Ref.~\cite{Li:2022dqa}.

In addition to the Super-K detector, this study also considers the JUNO detector~\cite{JUNO:2021vlw}. Because the electron recoil energy observed at JUNO is significantly larger than the binding energy, the corresponding ionization rate is similarly calculated using Eqs.\,(\ref{sk-crossing})--(\ref{range-sk-1}). 

\subsection{The recoil spectrum}

The electron recoil spectrum arising from CRDM scattering is calculated using Eq.\,(\ref{eq:ddrecoil}) for the PandaX-4T detector and Eq.\,(\ref{eq:vdrecoil}) for neutrino detectors, incorporating the DM flux following FSR. 

For the recoil rate calculations, we adopt total exposures of 198.9 tonne-days (Run0) and 363.3 tonne-days (Run1) for the PandaX-4T experiment, and 161.9 kiloton-years for Super-K. This normalization enables a direct spectral comparison with the data reported in Refs.\,\cite{PandaX:2024cic,Super-Kamiokande:2017dch}. For JUNO \cite{JUNO:2021vlw}, we calculate the recoil rates based on an assumed one-year exposure of its 20-kiloton liquid scintillator target. This scintillator composition corresponds to a total of $6.744 \times 10^{33}$ target electrons.\footnote{The mass fractions of carbon and hydrogen in the LAB-based liquid scintillator are approximately 88\% and 12\%, respectively.\cite{DayaBay:2015kir, JUNO:2021vlw}}
\begin{figure}[tb]
    \centering
    \includegraphics[width=0.32\textwidth]{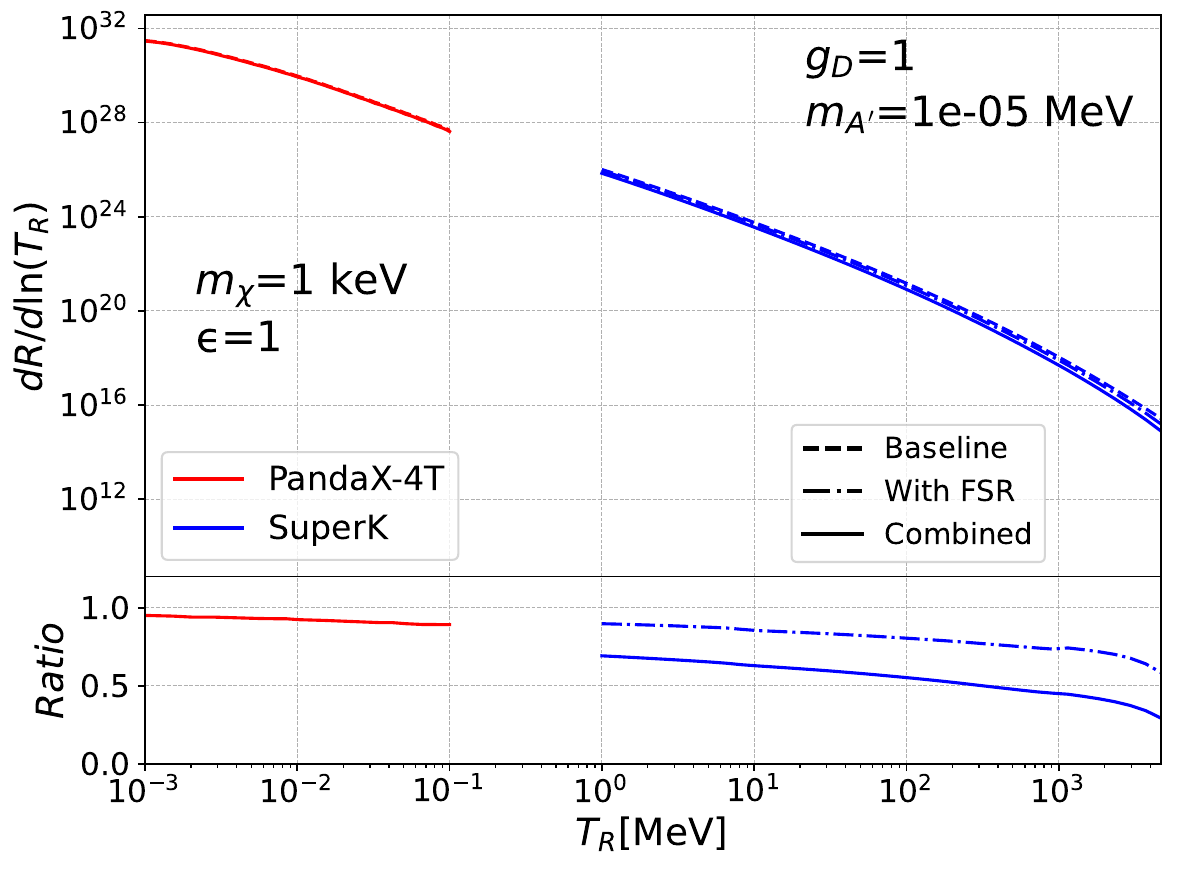}
    \includegraphics[width=0.32\textwidth]{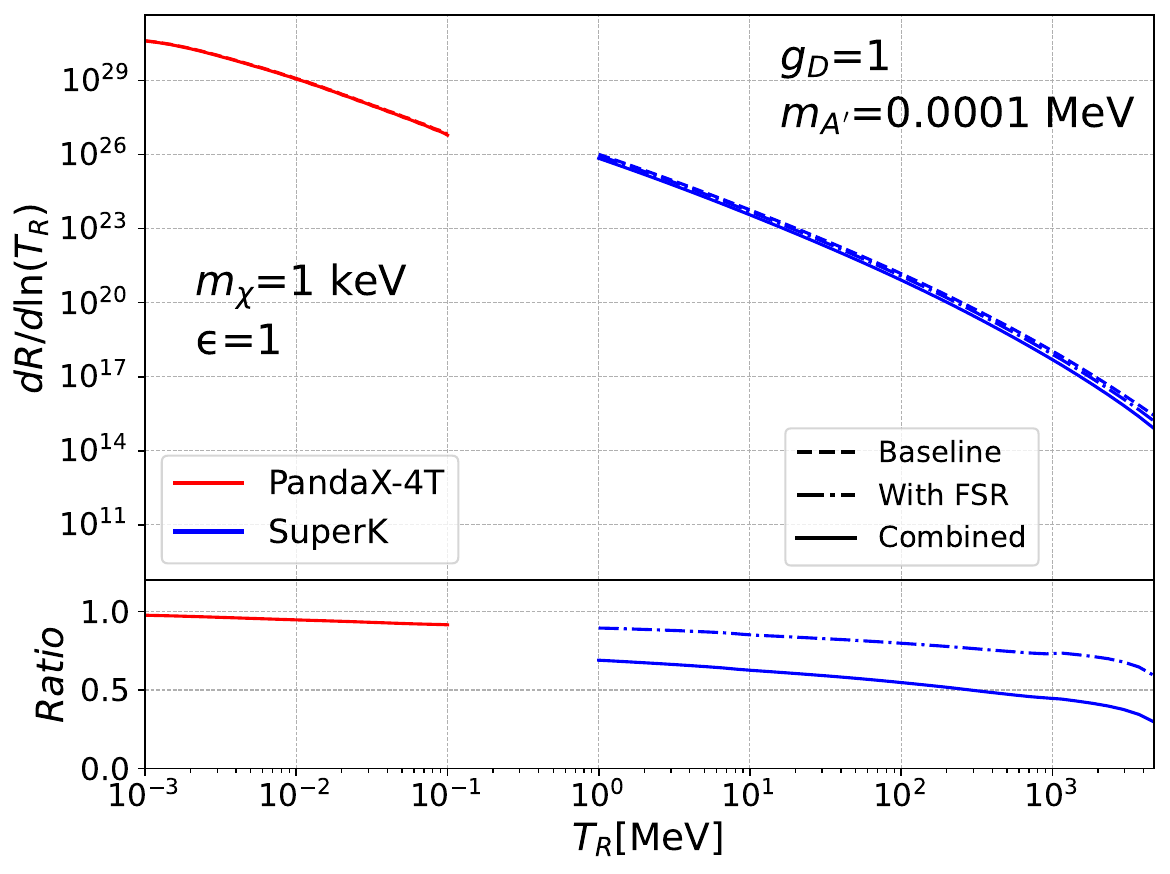}
    \includegraphics[width=0.32\textwidth]{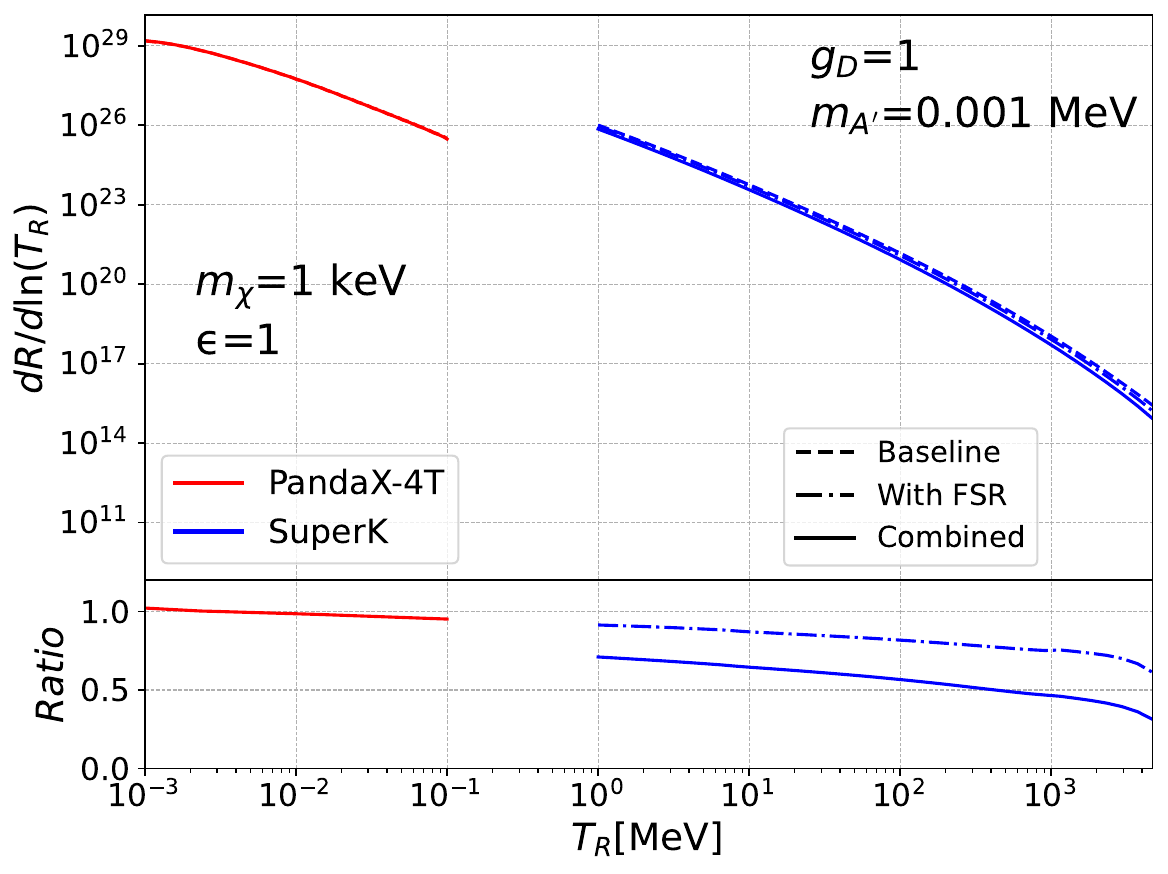}\\
    \includegraphics[width=0.32\textwidth]{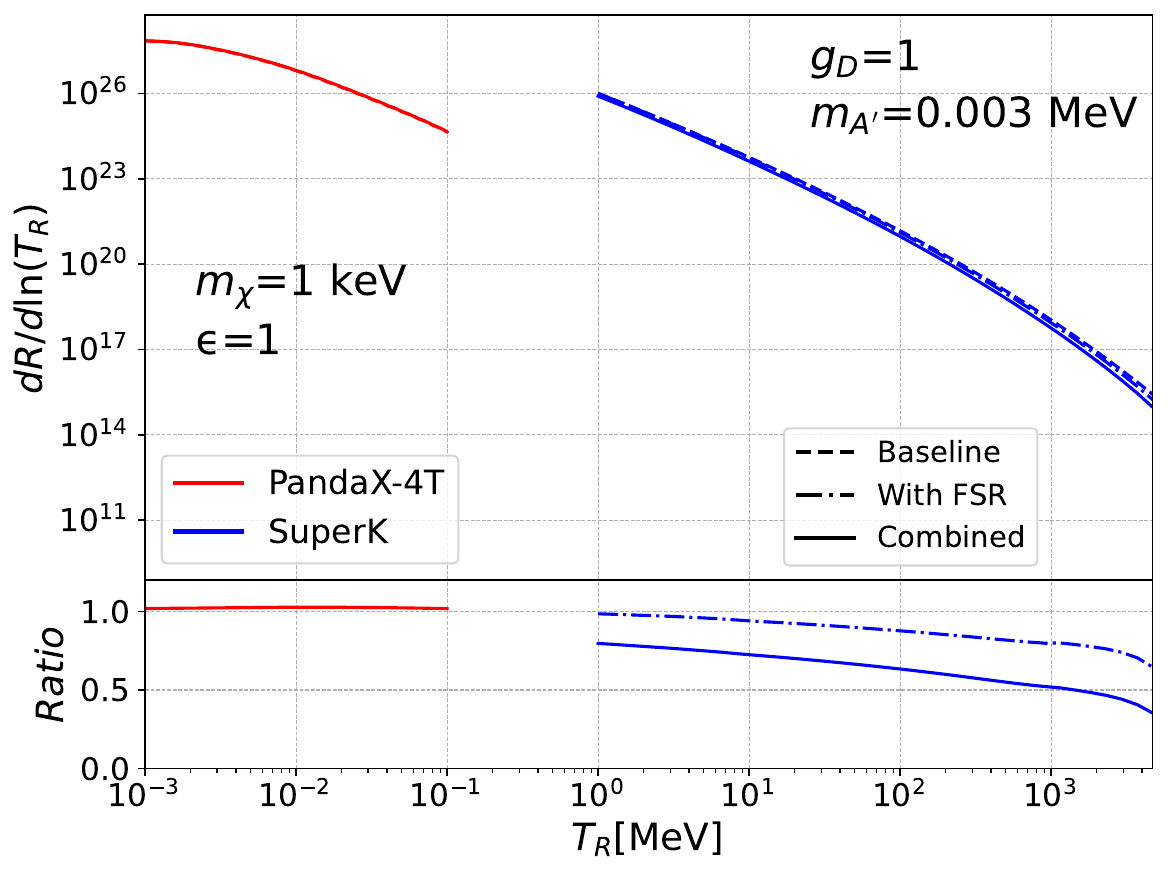}
    \includegraphics[width=0.32\textwidth]{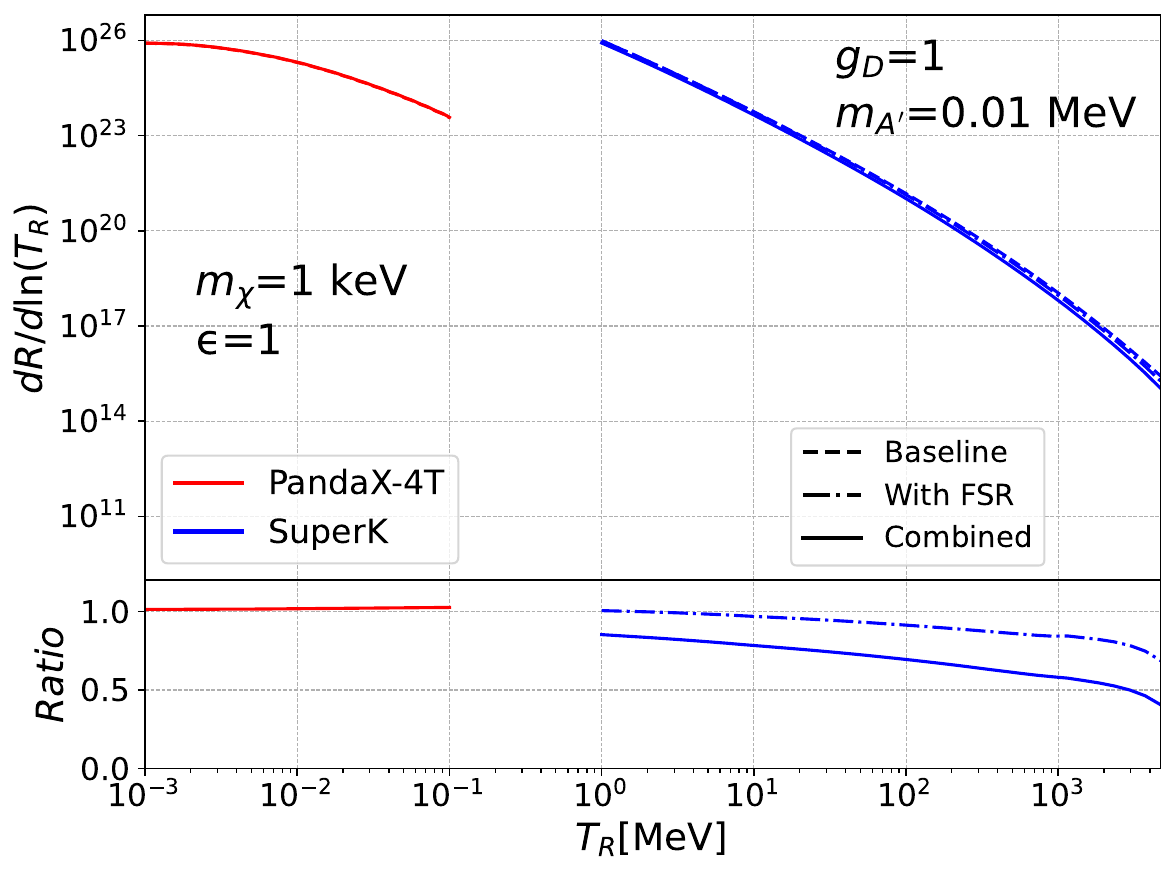}
    \includegraphics[width=0.32\textwidth]{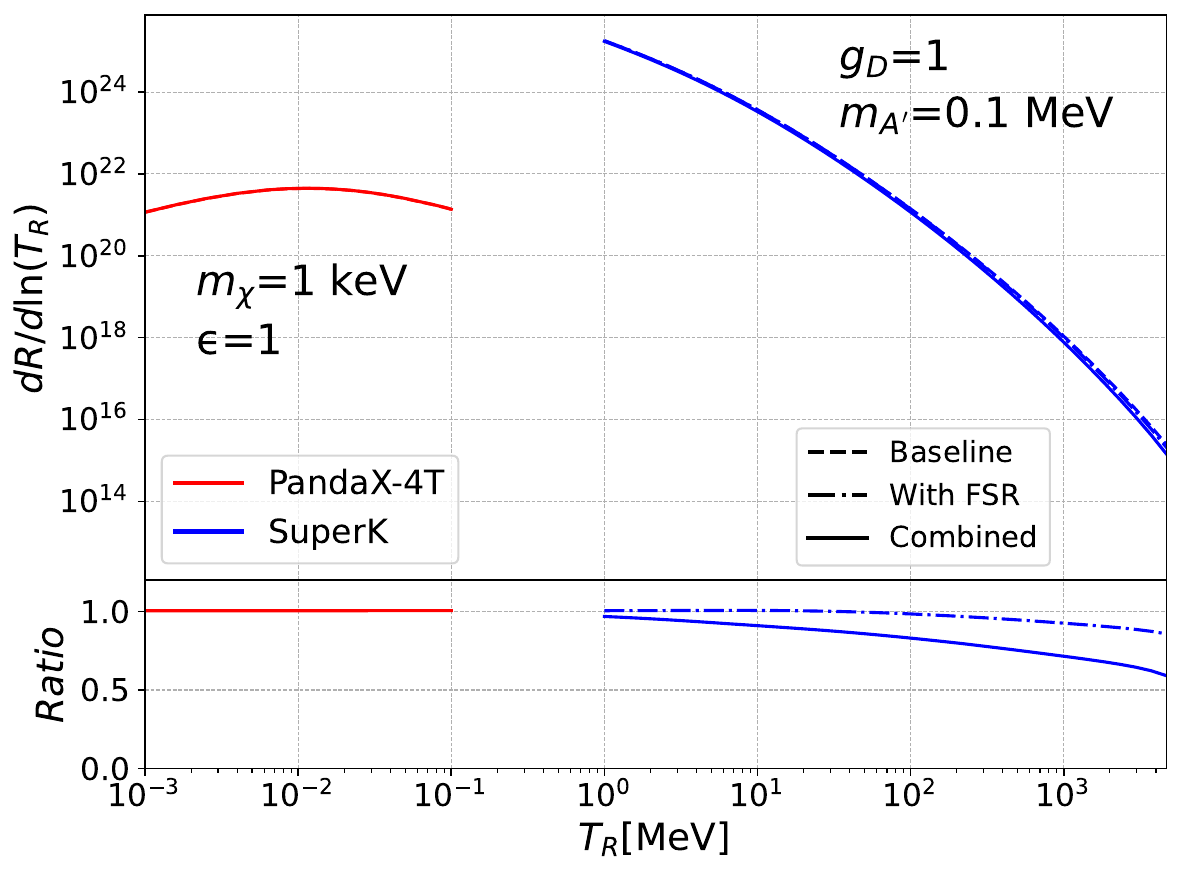}\\
    \caption{\label{fig:recoill1}The electron recoil rates in different detectors are shown for three cases: the Baseline (with neither FSR nor PDFs), FSR only, and the Combined scenario, which includes both FSR and PDF corrections. $g_D$ is set to 1, and the dark photon mass values are indicated in each plot. 
    The red and blue curves correspond to the rates for the PandaX-4T experiment with a 1.0 tonne-year exposure and the Super-K experiment with a data-taking period of 2628.1 days, respectively.}
\end{figure}

\begin{figure}[tb]
    \centering
    \includegraphics[width=0.32\textwidth]{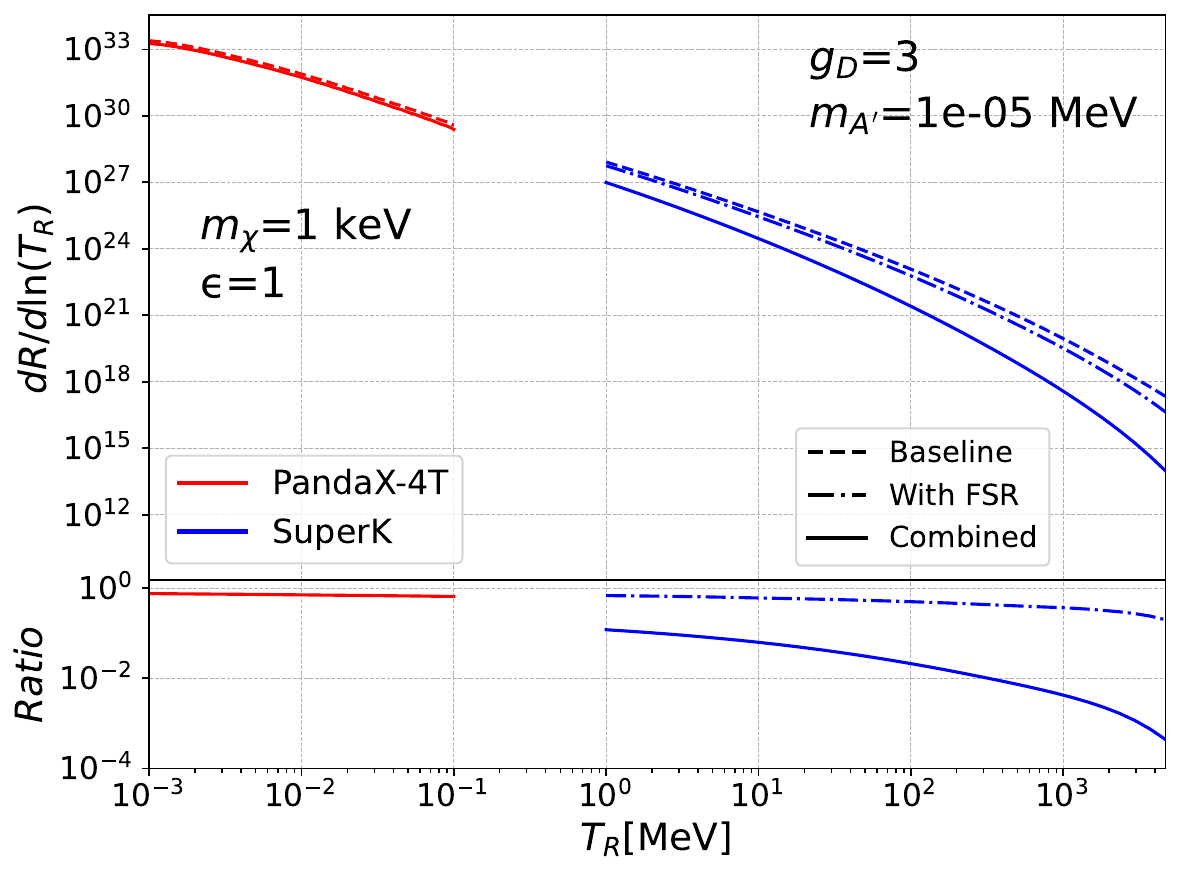}
    \includegraphics[width=0.32\textwidth]{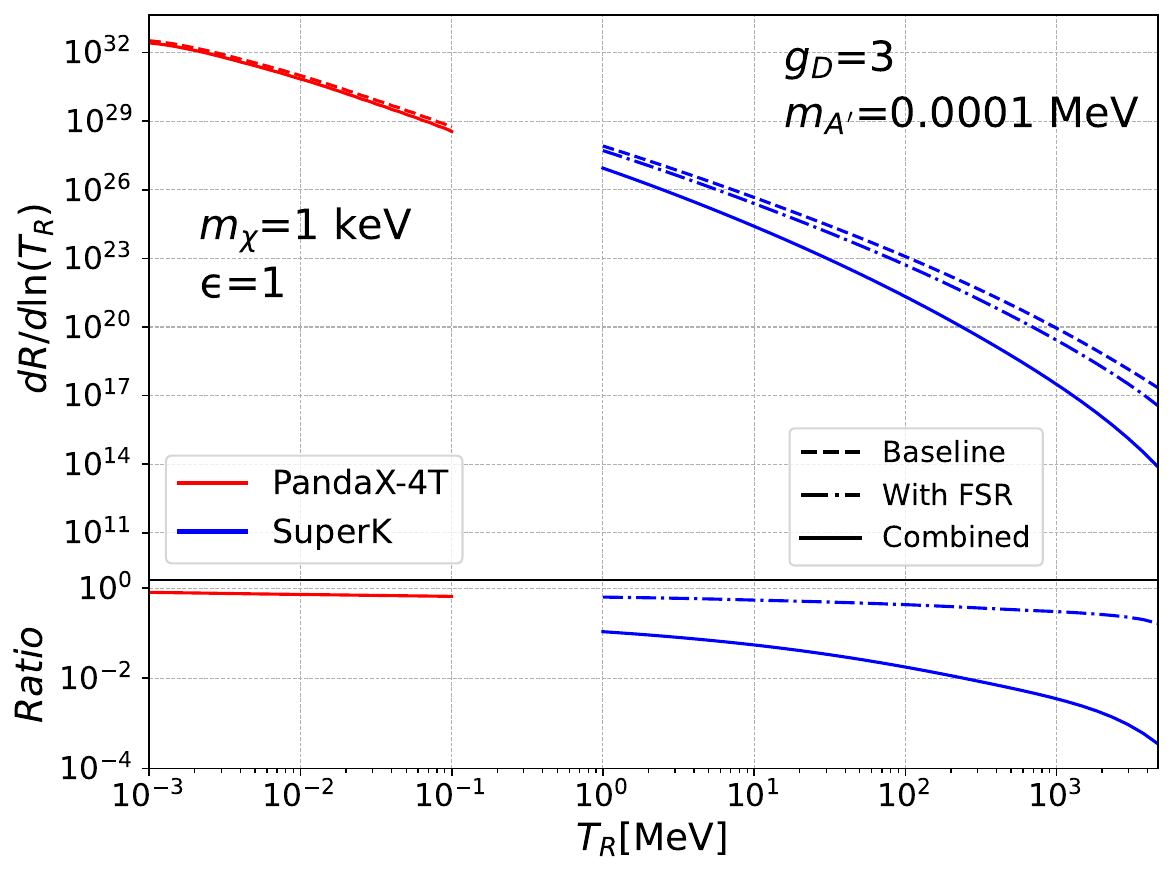}
    \includegraphics[width=0.32\textwidth]{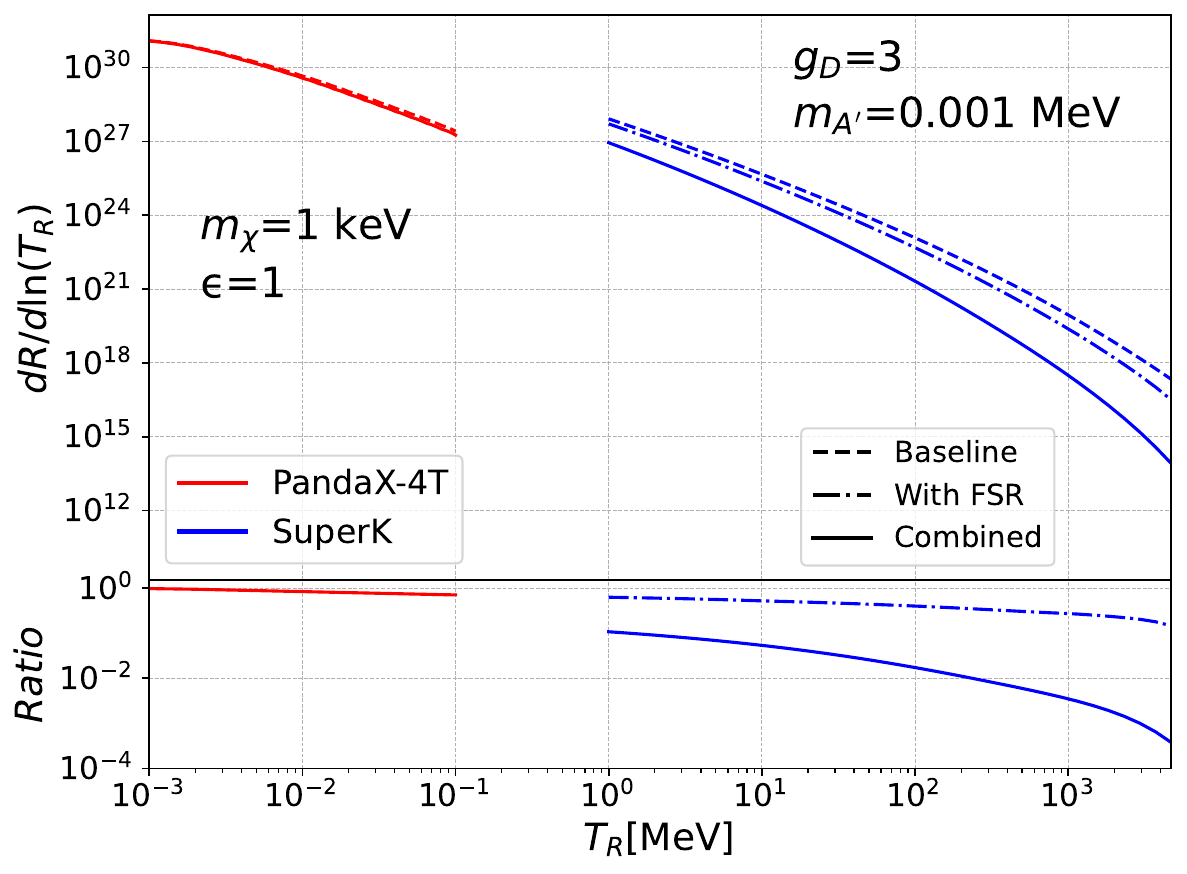}\\
    \includegraphics[width=0.32\textwidth]{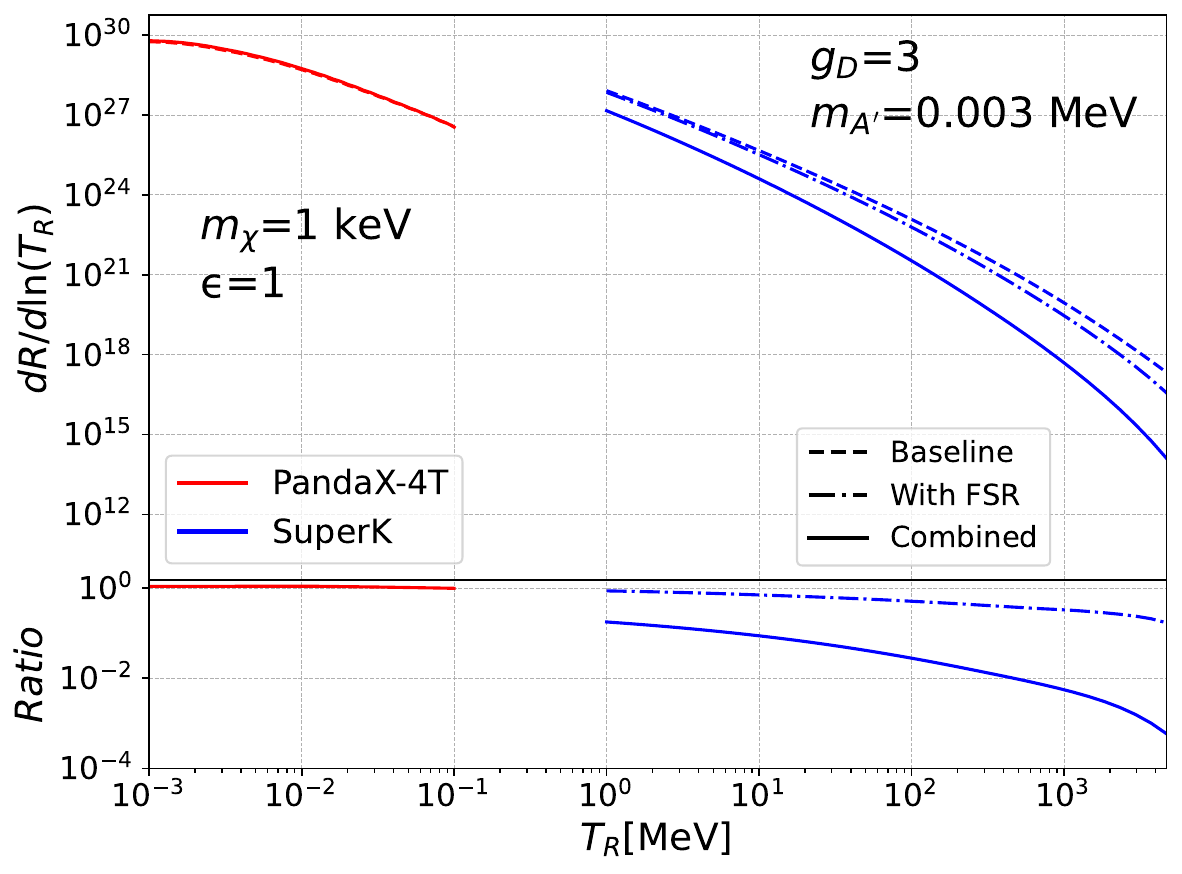}
    \includegraphics[width=0.32\textwidth]{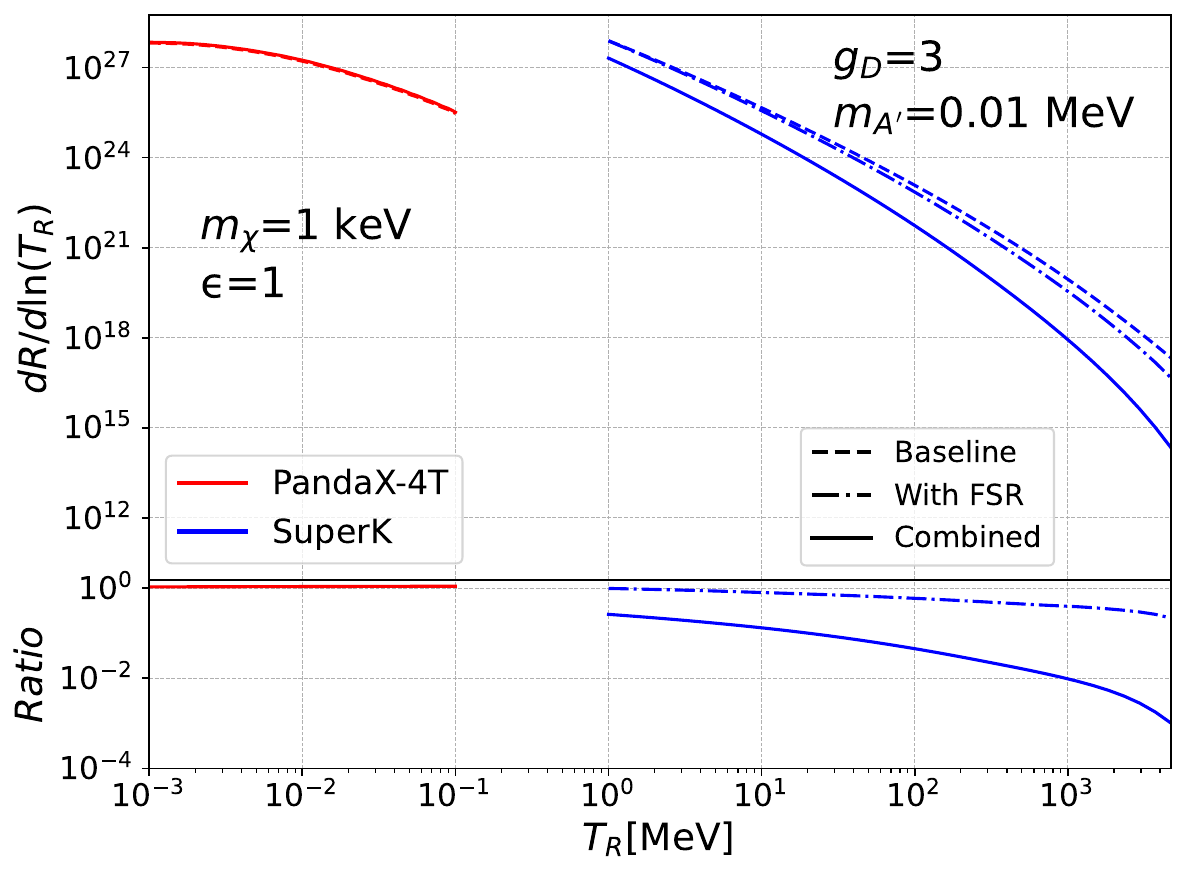}
    \includegraphics[width=0.32\textwidth]{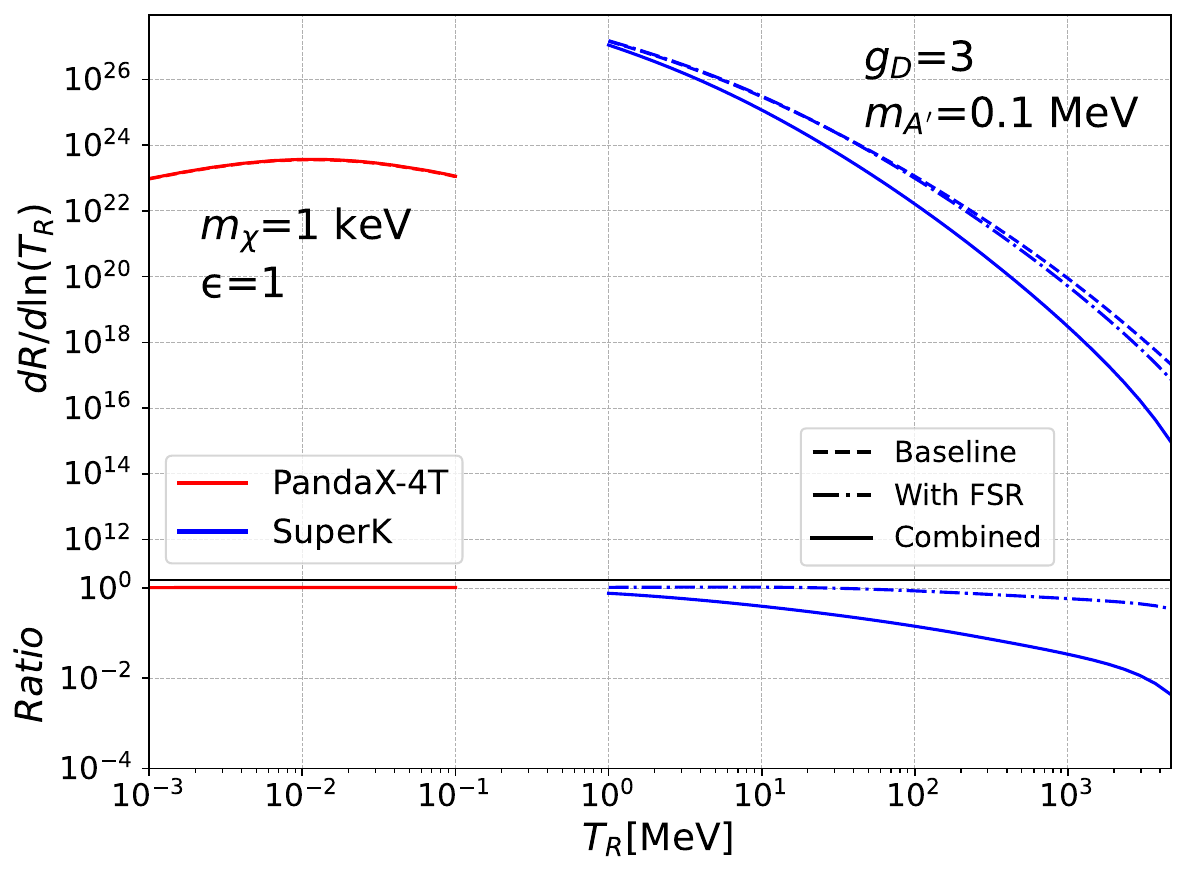}\\
    \caption{\label{fig:recoill3} The electron recoil rates in different detectors are shown for three cases: the Baseline (with neither FSR nor PDFs), FSR only, and the Combined scenario, which includes both FSR and PDF corrections. The coupling $g_D$ is set to 3, and the dark photon mass values are indicated in each plot. The red and blue curves correspond to the rates for the PandaX-4T experiment with a 1.0 tonne-year exposure and the Super-K experiment with a data-taking period of 2628.1 days, respectively.}
\end{figure}

Figures~\ref{fig:recoill1} and~\ref{fig:recoill3} depict the electron recoil rates in various detectors for CRDM with and without FSR, for $g_D$ values of 1 and 3, respectively.
For the PandaX-4T experiment, the recoil rate with a 1 tonne-year exposure is presented only in the low recoil energy region, as its measurements are limited to $E_R \in [0,30]$\,keV. The recoil rates for the Super-K neutrino detector are shown within the energy range $T_R \in [1,5\times10^4]$\,MeV. The recoil rates for the JUNO detector can be obtained by rescaling according to the number of target electrons and the exposure time.
The dependence of the recoil rates on the dark photon mass exhibits different behavior in the low and high recoil energy regions. At low recoil energies $T_R$, the rates are strongly suppressed by the dark photon mass $m_{A'}$, especially when the mass term dominates the propagators involved in the DM scattering off cosmic rays during acceleration and off target electrons during detection. This suppression becomes significantly weaker at higher $T_R$. As a result, neutrino detectors with higher energy thresholds may offer better sensitivity to CRDM if the dark photon mediator is relatively heavy.

The FSR effects on the recoil rates can be more easily seen through the dash-dotted curves in the lower panels of Figures\,\ref{fig:recoill1} and \ref{fig:recoill3}, which show the ratio of the recoil rate including FSR to the baseline (excluding both FSR and PDFs). These effects become more pronounced with stronger couplings and higher recoil energies. Moreover, they exhibit strong sensitivity to $m_{A'}$. For a lighter dark photon, the recoil rates are reduced across the entire $T_R$ range. For example, with $m_{A'}=10^{-4}$\,MeV and $g_D=1(3)$, the ratios of the recoil rates with and without FSR are 0.95 (0.72), 0.85 (0.54), and 0.8 (0.43) for $T_R=10$\,keV, 10\,MeV, and 100\,MeV, respectively, corresponding to the typical energy scales of the PandaX-4T, JUNO, and Super-K detectors. For a heavier dark photon, the recoil rates at PandaX-4T are enhanced at low $T_R$. As the mass increases, the enhancement region extends towards higher $T_R$, while the magnitude of the enhancement decreases. At high $T_R$, especially within the typical energy scale of Super-K, the recoil rates are significantly reduced. For example, with $m_{A'}=0.01$\,MeV and $g_D=1(3)$, the ratios of the recoil rates with and without FSR are 1.02 (1.08), 0.97 (0.8), and 0.92 (0.59) for $T_R=10$\,keV, 10\,MeV, and 100\,MeV, respectively. These FSR effects are consistent with the features of the CRDM flux discussed in the previous section.

When incorporating the effects of DM PDFs alongside FSR, the recoil rates undergo a further reduction, as illustrated by the solid curves in the lower panels. For a coupling of $g_D=1$, the magnitude of this suppression is of the same order as that induced by FSR. However, the effect becomes significantly more pronounced for $g_D=3$, with the suppression reaching $\mathcal{O}(0.01)$ in the large $T_R$ regime. This is primarily due to numerous splitting processes that substantially soften the incident DM spectrum. These features have been discussed in Ref.~\cite{Li:2022dqa}.

\section{Projected Sensitivities}
\label{sec5}

\subsection{Experimental bounds}\label{subsec:bound}

For the PandaX-4T experiment, the exclusion limits are derived using a $\chi^2$ analysis~\cite{Li:2022dqa,Li:2022jxo,Ghosh:2024dqw} of the recoiling electron spectrum,
\begin{align} 
  \chi^2
=
  \sum_i
\left(
  \frac {R_{\mathrm{\chi}}^i+R_{\mathrm{B_0}}^i-R_{\exp}^i}
        {\sigma_i}
\right)^2,
\end{align}
where $R_\mathrm{\chi}^i$, $R_{\mathrm{B_0}}^i$, and $R_{\exp}^i$ represent the theoretical prediction for the CRDM-induced recoil rate, background estimates, and observed recoil rates in the $i^{th}$ energy bin, respectively.
In the denominator, $\sigma_i$ represents the uncertainty associated with the observed data in the $i^{th}$ energy bin. 
The summation runs over all 60 energy bins across both the Run0 and Run1 datasets of the PandaX-4T experiment.
The observed data, background estimates, and associated uncertainties are taken from Ref.\,\cite{PandaX:2024cic}.
Since the test statistic follows a $\chi^2$ distribution with one degree of freedom, the exclusion regions corresponding to a 90\% confidence level (C.L.) are determined by applying the criterion $\Delta\chi^2=\chi^2-\chi^2_\mathrm{B_0} > 2.71$, where $\chi^2_\mathrm{B_0}$ is the $\chi^2$ value for the background-only case~\cite{Ghosh:2024dqw}.

The Super-K experiment conducted a boosted DM search using electron recoil events with kinetic energies $T_R > 100$\,MeV, analyzing data corresponding to a 161.9 kiloton-year exposure \cite{Super-Kamiokande:2017dch}.
Within the energy range $0.1\,\text{GeV} < T_R < 1.33\,\text{GeV}$, the total number of measured events $N_{\text{SK}}$ was 4042. 
Following the procedure outlined in Ref.\,\cite{Ema:2018bih}, a conservative upper limit on the DM recoil rate can be derived by imposing the condition:
\begin{align}
\xi \times R_\chi < N_{\text{SK}},
\end{align}
where $\xi=0.93$ represents the signal selection efficiency.
The recoil rate $R_\chi$ is calculated by integrating Eq.\,(\ref{eq:vdrecoil}) for $T_R$ above 100\,MeV, assuming a total of $N_e=7.5\times 10^{33}$ electrons and an exposure time of 2628.1 days. 

Although the JUNO detector achieves a recoil energy threshold as low as $\mathcal{O}(100)$\,keV, the neutrino background becomes negligible only for recoil energies $T_R \gtrsim 10$\,MeV.  
According to Ref.\,\cite{JUNO:2021vlw}, an estimated $\mathcal{O}(10)$ neutrino events are expected in the region $T_R>10$\,MeV for a 170 kiloton-year exposure. 
A conservative upper limit on the DM recoil rate is derived by integrating Eq.\,(\ref{eq:vdrecoil}) for $T_R$ above 10\,MeV and requiring this rate to be less than 10 events per year\footnote{A similar constraint is adopted in Ref.\,\cite{Dutta:2024kuj}}. 
\begin{figure}[tb]
    \centering
        \includegraphics[width=0.48\textwidth]{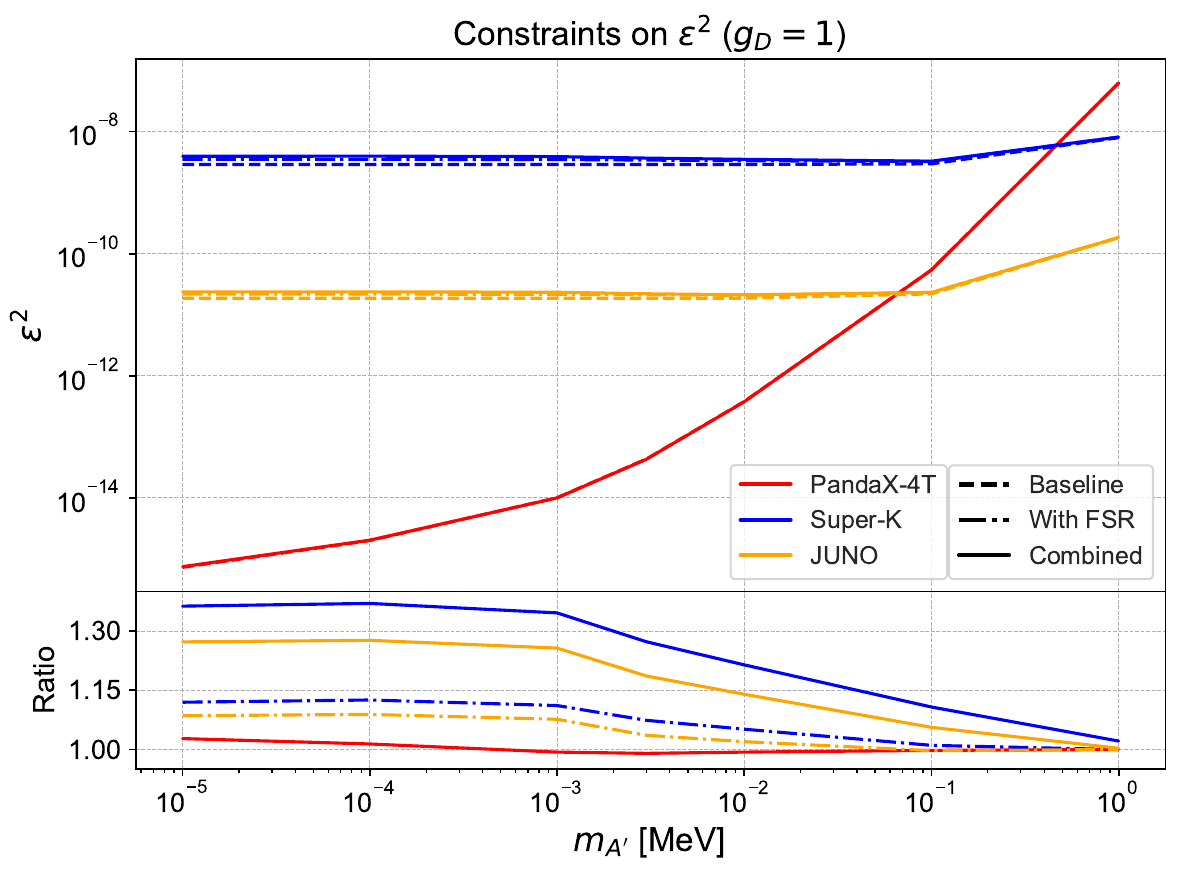} 
        \includegraphics[width=0.48\textwidth]{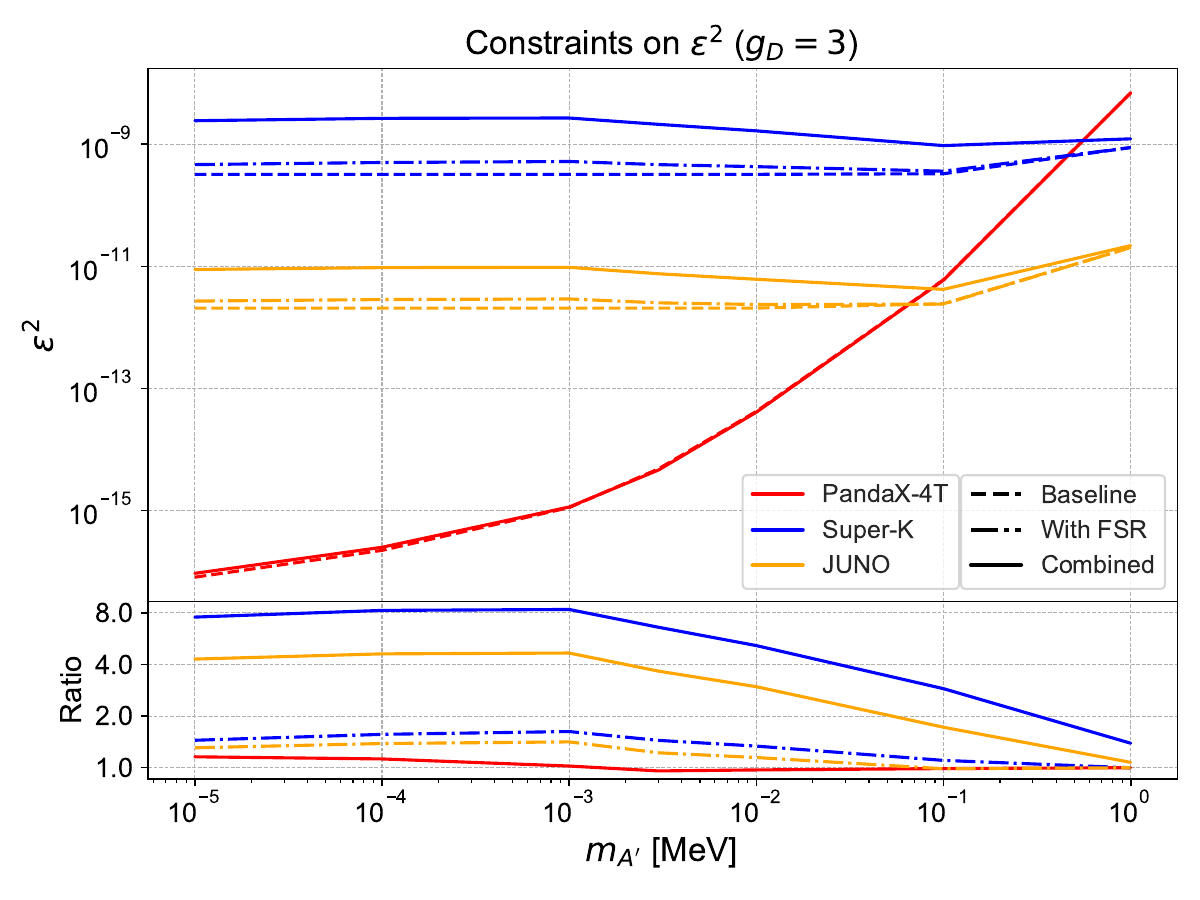} 
    \caption{\label{fig:bound} The constraints on the kinetic mixing parameter $\epsilon$ derived from the PandaX-4T, Super-K, and JUNO experiments are shown for $g_D=1$ (left) and $g_D=3$ (right). The dashed, dot-dashed and solid curves follow the same convention as in Figs. \ref{fig:recoill1} and \ref{fig:recoill3}. The lower panels show the ratios of the corresponding bounds on $\epsilon^2$ to the baseline bounds.}
\end{figure}

The resulting bounds on the kinetic mixing parameter $\epsilon$
for $g_D=1$ and $g_D=3$ are plotted in Figure~\ref{fig:bound}. 
The PandaX-4T experiment sets the most stringent bounds on a light dark photon compared to those from neutrino detectors. However, its sensitivity degrades dramatically with increasing dark photon mass, as the recoil rates at lower $T_R$ are suppressed by the mass term in the propagators. 
In contrast, at higher kinetic energies, both the CRDM flux and the corresponding recoil rates become less sensitive to $m_A'$. Consequently, Super-K and JUNO, which have higher energy thresholds, exhibit better sensitivity than PandaX-4T for heavier $m_A'$.

Given the FSR effects on the recoil rates, the ratio of the $\epsilon^2$ bounds with FSR to the baseline (with neither FSR nor PDFs) is shown in the lower panels of Figure~\ref{fig:bound}. The inclusion of FSR effects tends to relax the bounds by reducing the overall recoil rates. As this reduction is more significant at higher recoil energies, the impact of FSR is most pronounced for the Super-K experiment. 
As shown by the blue and yellow lines in the right panel, the bounds on $\epsilon^2$ are relaxed by factors of 1.6 and 1.4 at $m_A'=10^{-3}$\,MeV for Super-K and JUNO, respectively; these represent the most significant FSR effects across the entire dark photon mass parameter space\footnote{Since we did not sample more dark photon masses between 1\,keV and 3\,keV, it is difficult to pinpoint the precise mass value at which the FSR effect is most significant.}.
As explained in Secs.~\ref{subsec:kernel} and \ref{subsec:flux}, a heavier dark photon in the mass range $m_{A'} \lesssim 10^{-3}$\,MeV can efficiently soften the CRDM flux. As $m_{A'}$ increases further, the decay channel of the dark photon into a dark matter pair opens up, and the available phase space for shower evolution becomes increasingly restricted, resulting in less significant FSR effects. 
Finally, we note that FSR effects can also slightly strengthen the bounds of the PandaX-4T experiment, especially for $m_A'\gtrsim 3\times10^{-3}$\,MeV and $g_D=3$, where the CRDM flux is enhanced at $T_\chi\sim\mathcal{O}(10)$\,keV, as shown by the grey line in Figure~\ref{fig:fluxl1}.

Upon incorporating the DM PDF effects alongside FSR, the exclusion bounds are further relaxed, as illustrated by the solid curves in the lower panels. For a coupling of $g_D=1$, the magnitude of this relaxation is of the same order as that induced by FSR alone. However, the PDF effect becomes dominant for $g_D=3$.

Finally, we briefly comment on the effect of increasing the DM mass $m_\chi$. 
First, as shown in our previous work~\cite{Li:2022dqa}, increasing the DM mass reduces the pre-FSR flux values, steepens the decline of the flux curve, and decreases the sensitivity of direct detection experiments.\footnote{These behaviors are not significant when the dark photon is heavy, e.g., $\mathcal{O}(1)$\,MeV. However, we are not interested in the parameter space where both dark matter and dark photon are heavy, as the FSR effects are weak in this regime.} The second feature implies that a net enhancement of the flux after FSR becomes more difficult, as explained in Sec.~\ref{subsec:flux}. Overall, increasing the DM mass relative to the keV-scale DM adopted in this paper will suppress the FSR effects on the recoil spectrum at the DM detectors we consider. 
To elucidate this, we introduce a useful approximation: for a rapidly falling CRDM flux spectrum, the FSR effects on the recoil rate at $T_R$ can be effectively understood through the corresponding effects on the CRDM flux at $T_\chi$ of the same order as $T_R$. This is because the recoil rate of electrons with kinetic energy $T_R$ is dominated by scattering processes where the initial CRDM carries a kinetic energy $T_\chi$ of the same order as $T_R$, whereas contributions from more energetic CRDM are strongly suppressed by the flux. With this approximation, it becomes clear why the FSR effects at DM detectors are suppressed for DM masses heavier than our baseline keV-scale choice. First, $T_{\chi,\text{min}}^{\text{FSR}}$ increases with $m_\chi$ and can easily exceed the recoil energy scale of the PandaX-4T experiment ($\mathcal{O}(10)$\,keV). For instance, $m_\chi=1$\,MeV leads to $T_{\chi,\text{min}}^{\text{FSR}}\gtrsim 0.5$\,MeV.  
Therefore, FSR effects on the CRDM flux at the $\mathcal{O}(10)$\,keV $T_\chi$ scale are negligible\footnote{This is because DM in this low-energy band ($T_\chi\sim\mathcal{O}(10)$\,keV) is kinematically forbidden from participating in FSR, and the number of new particles emitted into this band by high-energy ($T_\chi>0.5$\,MeV) DM is negligible compared to the original population. Such behavior was demonstrated in Sec.~\ref{subsec:flux}.}, resulting in negligible FSR effects at the PandaX-4T experiment.
Second, the hierarchy between the maximum evolution variable $Q_\text{max}$ (proportional to $\sqrt{m_\chi}$, as shown in Sec.~\ref{subsec:FSR}) and $m_\chi$ gradually diminishes as $m_\chi$ increases. This hierarchy is the key framework for analyzing FSR. 
For $T_\chi$ at the $\mathcal{O}(10)$\,MeV scale---the typical recoil energy scale at neutrino detectors---the hierarchy becomes even weaker. For instance, with $m_\chi=1$\,MeV and $T_\chi=10$\,MeV, the ratio $Q_\text{max}/m_\chi=4.47$, whereas this ratio is 141 for $m_\chi=1$\,keV (our baseline choice). The limited evolution space implies that MeV-scale DM exhibits much weaker FSR effects (relative to keV-scale DM) on the CRDM flux at $T_\chi\sim \mathcal{O}(10)$\,MeV, consequently leading to significantly weaker FSR effects on the recoil spectrum at neutrino detectors.

\subsection{Constraints from the dark matter self-interaction} \label{sec:sidm}

A light mediator and substantial coupling lead to significant DM self-scattering. The corresponding cross section is constrained by observations, including the Bullet Cluster \cite{Markevitch:2003at,Clowe:2003tk,Randall:2008ppe, Kahlhoefer:2013dca}, and cosmological simulations of self-interacting DM on galactic and galaxy cluster scales \cite{Rocha:2012jg,Peter:2012jh}. 

In the present framework, DM self-interactions are mediated by the dark photon. In the non-relativistic limit, scattering between $\chi$ and $\bar{\chi}$ is governed by the attractive Yukawa potential $V(r) \equiv - \alpha^\prime e^{- m_{A'} r} / r$, derived from the coupling term in Eq.\,(\ref{xxa}) with $\alpha^\prime \equiv g_D^2 / 4 \pi$. The corresponding scattering amplitude is given by
\begin{align}  
  f(\theta)
=
  \frac 1 k
  \sum_{l=0}^\infty (2l+1)e^{i\delta_l}P_l(\cos\theta)\sin\delta_l.  
\end{align}  
Here, $\delta_l$ denotes the phase shift for the $l$-th partial wave, obtained by solving the Schrödinger equation with the Yukawa potential $V(r)$. The momentum parameter $k$ is defined as  
$k \equiv m_\chi v / 2$,  
where $v$ denotes the relative velocity between $\chi$ and $\bar{\chi}$.  
Because the total scattering cross section $\sigma=\int\left|f(\theta)\right|^2d\Omega$ diverges, we instead characterize $\chi$-$\bar{\chi}$ scattering using the transfer cross section $\sigma_\text{T}$ and the viscosity cross section $\sigma_\text{V}$, defined as \cite{PhysRevA.60.2118}
\begin{align}  
    & \sigma_\text{T}=\int d\Omega(1-\cos\theta)\frac{d\sigma}{d\Omega}=  
    \frac{4\pi}{k^2}\sum_{l=0}^\infty(l+1)\sin^2(\delta_{l+1}-\delta_l)~,\\ 
    & \sigma_\text{V}=\int d\Omega\sin^2\theta\frac{d\sigma}{d\Omega}=\frac{4\pi}{k^2}\sum_{l=0}^\infty\frac{(l+1)(l+2)}{2l+3}\sin^2(\delta_{l+2}-\delta_l)~.  
\end{align}  

Following the convention established in Refs.\,\cite{Buckley:2009in,Tulin:2013teo}, we introduce the dimensionless parameters $a \equiv v / 2 \alpha^\prime$, $b \equiv \alpha^\prime m_\chi / m_{A^\prime}$, and $t \equiv a b$. For the parameter space of interest in this work ($g_D\sim\mathcal{O}(1)$, $v\sim 1000\,\text{km/s}$, $m_\chi=1\,\text{keV}$, $m_{A^\prime}/\text{MeV}\in[10^{-5},1]$), the condition $t<1$ holds. This implies that the $s$-wave phase shift is significantly larger than those of higher partial waves ($|\delta_0|\gg |\delta_{l}|$ for $l>0$), and consequently, $s$-wave scattering dominates the interaction. Under the $\text{Hulth\'{e}n}$ approximation, the cross sections can be expressed in a simplified form as \cite{Tulin:2013teo,Colquhoun:2020adl}
\begin{align}  
    \sigma_\text{T}\approx\frac{3}{2}\sigma_\text{V}\approx\frac{4\pi}{k^2}  
    \sin^2\left(\delta_0^\text{Hulth\'{e}n}\right)\label{sigT} 
\end{align}  
with the corresponding $s$-wave phase shift given by~\cite{Tulin:2013teo}:
\begin{align}  
\delta_0^\text{Hulth\'{e}n}=\text{arg}\left(\frac{i\Gamma(\lambda_++\lambda_--2)}{\Gamma(\lambda_+)\Gamma(\lambda_-)}\right)~, 
\end{align}  
where $\lambda_\pm \equiv 1+iac\pm\sqrt{c-a^2c^2}$ and $c\approx b/1.6$.

\begin{figure}[tb]
    \centering
        \includegraphics[width=0.4\textwidth]{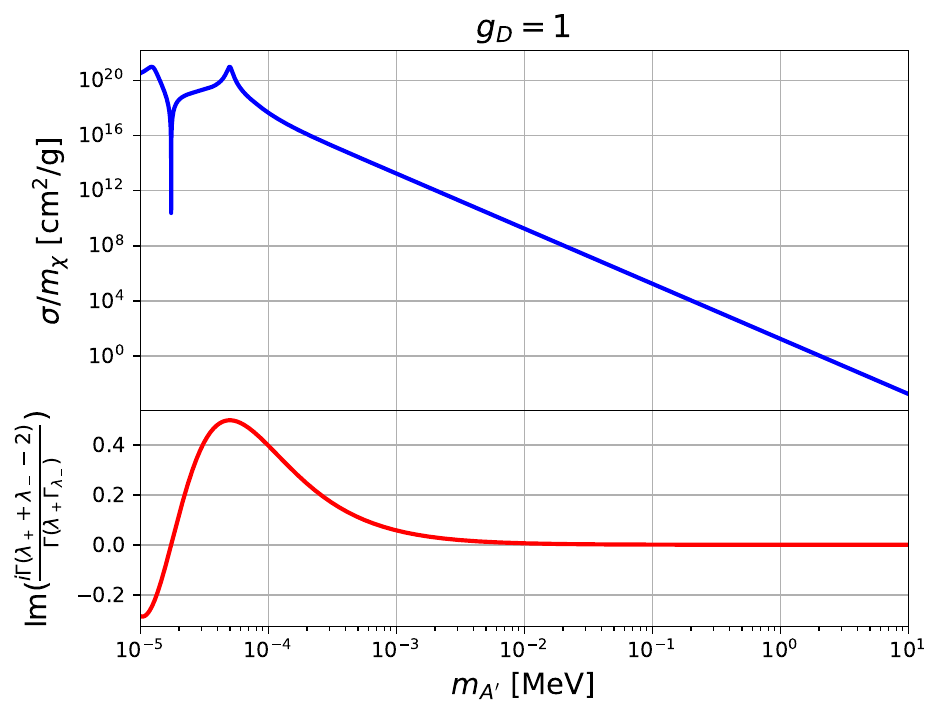} 
        \includegraphics[width=0.4\textwidth]{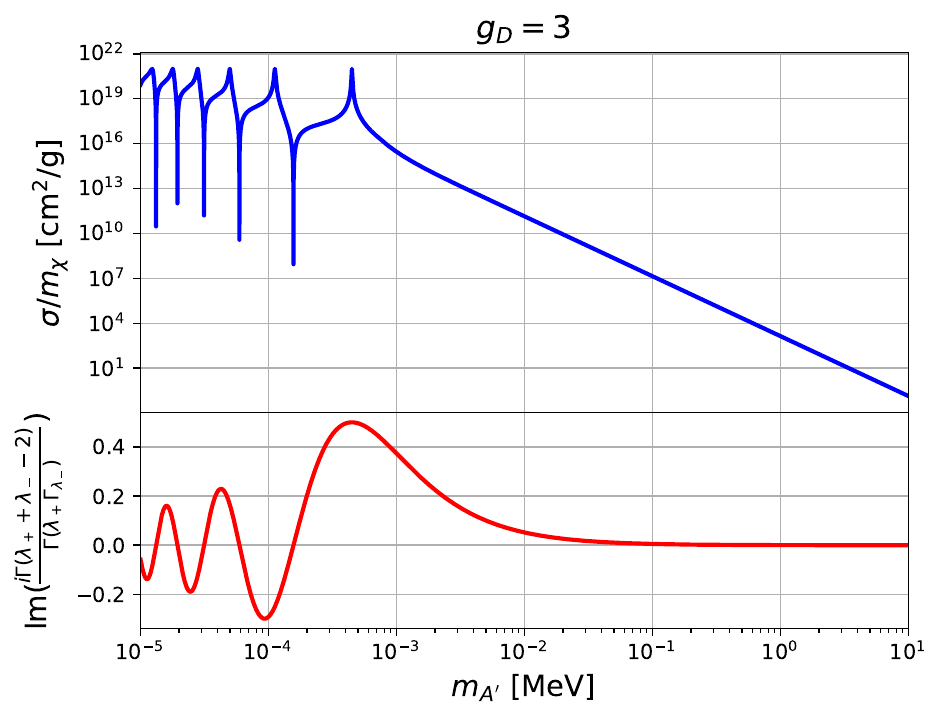} 
\caption{\label{fig:sid} The $\sigma_\text{T}/m_\chi$ ratio obtained from Eq.\,(\ref{sigT}) (upper panels) and the imaginary part of \( \frac{i\Gamma(\lambda_++\lambda_--2)}{\Gamma(\lambda_+)\Gamma(\lambda_-)} \) (lower panels) for $g_D=1$ and $g_D=3$ (left and right panels, respectively). The relative velocity between the dark matter particles is set to $v=1000~\text{km/s}$.}
\end{figure}

The constraints derived from observations of galaxy clusters, which correspond to a characteristic velocity of $v=1000~\text{km/s}$, require the self-interaction cross section to satisfy $\sigma_\text{T}/m_\chi\lesssim 1~\text{cm}^2/\text{g}$ \cite{Tulin:2013teo,Laha:2013gva}. In Figure\,\ref{fig:sid}, we present $\sigma_\text{T}/m_\chi$ as calculated from Eq.\,(\ref{sigT}), alongside the imaginary part of $i \Gamma(\lambda_++\lambda_--2) / \Gamma(\lambda_+)\Gamma(\lambda_-)$. These quantities are evaluated for a relative velocity of $v=1000~\text{km/s}$ between dark matter particles, considering both coupling values $g_D=1$ and $g_D=3$. 
We find that in addition to the regime where $m_{A^\prime}$ exceeds the MeV scale (and thus satisfies $\sigma_\text{T}/m_\chi\lesssim 1\,\text{cm}^2/\text{g}$), these constraints are also met at specific resonant points. 
These points correspond to the zeros of the imaginary part of $i \Gamma(\lambda_++\lambda_--2) / \Gamma(\lambda_+)\Gamma(\lambda_-)$, which enforce a vanishing phase shift, $\delta_0^\text{Hulth\'{e}n}=0$, resulting in a suppressed cross section, $\sigma_\text{T}/m_\chi \sim 0$. 
For a given $m_A^\prime \gtrsim 10^{-5}~\text{MeV}$, the number of these zeros increases as $g_D$ and $m_\chi$ increase. 
Numerical calculations yield the maximum $m_{A^\prime}$ for which these resonances can occur:
\begin{align}  
  m_{A^\prime}
\approx
  1.7\times10^{-5}\times g_D^2\times\left(\frac{m_\chi}{\text{keV}}\right)\,\text{MeV}.  
\end{align}
By systematically adjusting $g_D$ and $m_\chi$, the constraint $\sigma_\text{T}/m_\chi\lesssim 1\,\text{cm}^2/\text{g}$ can be maintained across multiple orders of magnitude in $m_{A^\prime}$ values above $10^{-5}\,\text{MeV}$. 

However, achieving $\delta_0^\text{Hulth\'{e}n}=0$ by tuning the parameters ($g_D$, $m_\chi$, $m_{A^\prime}$, $v$) to satisfy $\sigma_\text{T}/m_\chi\lesssim 1\,\text{cm}^2/\text{g}$ inevitably introduces a fine-tuning problem. Our numerical calculations show that, for the parameter choices in Figure~\ref{fig:sid} that satisfy $\delta_0^\text{Hulth\'{e}n}=0$, even a $1\%$ variation in $v$ causes $\sigma_\text{T}/m_\chi$ to increase sharply to at least $10^6~\text{cm}^2/\text{g}$ (for fixed $m_\chi$ and $g_D$, a larger $m_{A^\prime}$ yields a smaller enhancement). Moreover, under the more realistic assumption that the relative velocity $v$ between DM particles follows a distribution $f(v)$ rather than remaining fixed, such parameter tuning fails to achieve $\sigma_\text{T}/m_\chi\lesssim 1\,\text{cm}^2/\text{g}$. This conclusion also applies to a wider parameter space of CRDM scenarios~\cite{Li:2022dqa,Xu:2024iny}\footnote{While many studies adopt a total cross-section approach without specifying the propagator details, implementing a consistent UV completion that evades current mediator exclusion limits remains non-trivial.}.

This motivates an extension of the model that simultaneously suppresses DM self-interactions and leaves the dark shower kinematics unaltered compared to the model presented here. In a more general framework, the dark shower phenomenology can be disentangled from DM self-interaction constraints by invoking multi-component or inelastic dark matter scenarios~\cite{Tucker-Smith:2001myb,Ahmed:2017dbb,Ko:2019wxq,Baek:2020owl,Wang:2025xoq}. In specific U(1) extended models, the vector mediator couples exclusively to distinct mass eigenstates (denoted as $\chi_1$ and $\chi_2$) via off-diagonal currents, forbidding the tree-level diagonal coupling $A^\prime \chi_1 \chi_1$. Since the relic abundance is dominated by the lighter state $\chi_1$, tree-level elastic self-scattering is eliminated, thereby satisfying SIDM constraints. Meanwhile, the kinematics of the dark shower remain unaltered, provided the mass splitting between the eigenstates is negligible compared to the high energy of the cosmic rays and the DM-electron scattering energy scale. Consequently, the main conclusions of this work remain robust.

We note that higher-order Feynman diagrams could reintroduce DM self-interactions. However, these contributions depend strongly on the specific Lorentz structure of the couplings (e.g., vector vs. axial-vector) and the precise mass spectrum. We leave a dedicated analysis of these complex loop-level effects for future work.

\section{Conclusion and Discussion}
\label{sec6}

In this work, we systematically investigate dark parton shower effects in the direct detection of CRDM within a dark photon-mediated fermionic DM model. By developing a Monte Carlo framework that incorporates Sudakov form factors and kinematic dipole recoil schemes, we simulate the evolution of DM energy spectra under dark photon splitting and quantify the impact of FSR on experimental sensitivities. 

By examining the FSR evolution kernel, which governs the energy redistribution of DM particles, we elucidate the primary characteristics of the showering process. Specifically, we highlight the resulting degradation of the DM energy and identify the minimum initial DM energy required for FSR to be kinematically allowed.
FSR depletes the flux of high-energy DM particles while generating an excess of secondary DM particles at lower energies. The net outcome depends strongly on the shape of the pre-FSR flux and the particle masses. 
For instance, with $2m_\chi \lesssim m_{A^\prime} \lesssim 10^{-2}$\,MeV
and $g_D=3$, the CRDM flux can be enhanced by a factor of up to 1.12 in the $\mathcal{O}(10^{-2} \sim 1)$\,MeV energy range. 
Conversely, for a lighter mediator with $m_{A^\prime} \lesssim 10^{-3}$\,MeV and the same coupling, FSR reduces the DM flux at $\sim 100$\,MeV by more than 50\%. 

The modified CRDM flux directly impacts the electron recoil rates in both DM and neutrino experiments. In high-threshold detectors such as Super-K and JUNO, the dominant effect of FSR is the depletion of the high-energy DM flux. This suppresses the signal rate, systematically weakening the experimental constraints. For example, at $m_{A^\prime} = 10^{-3}$\,MeV and $g_D=3$, the bounds on $\epsilon^2$ are relaxed by factors of 1.6
and 1.4 at Super-K and JUNO, respectively. 
Conversely, for low-threshold experiments such as PandaX-4T, a slight signal enhancement is predicted for $m_{A^\prime} \sim 3 \times 10^{-3}$\,MeV. This occurs because abundantly radiated dark photons decay back into DM pairs, replenishing the DM flux in the $\mathcal{O}(10)$\,keV energy range. We note that FSR effects become negligible if the DM is much heavier than the keV scale, e.g., $\mathcal{O}$(1) MeV, due to the limited phase space for evolution.

Incorporating DM PDF effects alongside FSR further relaxes the recoil rates and corresponding exclusion bounds. This effect is particularly pronounced for larger couplings, where the DM PDFs significantly soften the incident DM spectrum. This behavior is consistent with the findings in Ref.~\cite{Li:2022dqa}.

We also examine constraints from DM self-interactions. By introducing a multi-component dark matter scenario, the SIDM constraints can be naturally avoided, and the dark shower phenomenology remains unchanged compared to the model studied in this paper, provided that the mass splitting between different dark matter components is negligible relative to both the high energy of cosmic rays and the DM-electron scattering energy scale. Consequently, the main conclusions of this work remain robust.
Finally, we briefly comment on other potential constraints and future directions relevant to this scenario.

As summarized in Ref.\,\cite{Fabbrichesi:2020wbt}, various experiments place strong constraints on the kinetic mixing parameter $\epsilon$ within our parameter region of interest, \textit{i.e.} $10^{-5}\,\text{MeV}<m_{A'}<1$\,MeV. The most stringent constraints in this mass range arise from the emission of dark photons from stars~\cite{An:2013yfc, Redondo:2013lna}, such as the Sun, horizontal-branch stars, and red giants. The absence of anomalous energy loss in these stars, as well as the detection of solar dark photons by experiments such as XENON10~\cite{An:2013yua}, CAST~\cite{Redondo:2008aa}, and SHiP~\cite{Schwarz:2015lqa}, imposes stringent bounds on the kinetic mixing parameter. However, these studies generally neglect self-interactions within the dark sector when modeling dark photon emission from stars. As pointed out in Ref.\,\cite{Sung:2021swd, Ganguly:2006ki}, the presence of self-interactions within dark sectors can significantly reduce the mean free path of dark species, effectively trapping them inside stars and preventing their free escape. Therefore, this mechanism can considerably suppress radiative transfer. Using the same dark photon model considered in our work, Ref.\,\cite{Sung:2021swd} demonstrates that even a small dark sector coupling ($\alpha_D \ll 1$) can lead to efficient self-trapping of dark particles in proto-neutron stars. Ref.\,\cite{Ganguly:2006ki} also shows that solar constraints on the pseudoscalar-photon coupling can be evaded by studying the $\phi^4$ self-interaction term in a pseudoscalar model.  

In addition to stellar cooling constraints, beam dump experiments such as NA64~\cite{NA64:2023wbi, Banerjee:2019pds, Ge:2025aui} also provide limits on the mixing parameter from searches for missing-energy events. However, these experiments are primarily sensitive to masses $m_A'\gtrsim 1$\,MeV. At the boundary of this region, $m_A'=1$\,MeV, the 90\% C.L. bound is slightly stronger than the JUNO limit derived in this work for $g_D=1$, but weaker than that for $g_D=3$. Although bounds extrapolated into the region $m_{A'}\lesssim 1$\,MeV are presented in Ref.\,\cite{Fabbrichesi:2020wbt}, they are not competitive with the constraints we derive from the JUNO experiment.

\begin{acknowledgments}

This work was supported in part by the Natural Science Foundation of Sichuan Province under grants No. 2026NSFSC0034, by the National Natural Science Foundation of China under grants No. 11905149 and No. 12505121,  by the Joint Fund of Henan Province Science and Technology R$\&$D Program (Project No. 245200810077), by the Startup Research Fund of Henan Academy
 of Sciences (Project No. 20251820001), and by the Scientific and Technological Research Project of Henan Academy of Sciences (Project No. 20262320001).
SFG is supported by the National Natural Science
Foundation of China (12375101, 12425506, 12090060 and 12090064) and the SJTU Double First
Class start-up fund (WF220442604).
SFG is also an affiliate member of Kavli IPMU, University of Tokyo.
C.Z. acknowledges the SinoGerman (CSC-DAAD) Postdoc Scholarship Program,
2023 (57678375).

\end{acknowledgments}

\bibliographystyle{utphysGe}
\bibliography{BoostDM}
\end{document}